\documentclass[%
onecolumn,
superscriptaddress,
 amsmath,amssymb,
 aps,
]{revtex4-2}

\usepackage[colorlinks,citecolor=blue,linkcolor=blue,urlcolor=blue]{hyperref}
\usepackage{graphicx}
\usepackage{dcolumn}
\usepackage{bm}
\usepackage{subcaption}
\usepackage{upgreek}
\usepackage{xcolor}
\usepackage{tabularx}
\usepackage{booktabs}

\begin{document}

\title{A stochastic model for the hydrodynamic force in Euler--Lagrange simulations of particle-laden flows}

\author{Aaron M. Lattanzi}
\email[]{alattanz@umich.edu}
\affiliation{University of Michigan, Department of Mechanical Engineering, Ann Arbor, MI}

\author{Vahid Tavanashad}
\affiliation{Iowa State University, Department of Mechanical Engineering, Ames, IA}

\author{Shankar Subramaniam}
\affiliation{Iowa State University, Department of Mechanical Engineering, Ames, IA}

\author{Jesse Capecelatro}
\affiliation{University of Michigan, Department of Mechanical Engineering, Ann Arbor, MI}
\affiliation{University of Michigan, Department of Aerospace Engineering, Ann Arbor, MI}

\date{\today}

\begin{abstract}
Standard Eulerian--Lagrangian (EL) methods generally employ drag force models that only represent the mean hydrodynamic force acting upon a particle-laden suspension. Consequently, higher-order drag force statistics, arising from neighbor-induced flow perturbations, are not accounted for; with implications on predictions for particle velocity variance and dispersion. We develop a force Langevin (FL) model that treats neighbor-induced drag fluctuations as a stochastic force within an EL framework. The stochastic drag force follows an Ornstein-Uhlenbeck process and requires closure of the integral time scale for the fluctuating hydrodynamic force and the standard deviation in drag. The former is closed using the mean-free time between successive collisions, derived from the kinetic theory of non-uniform gases. For the latter, particle-resolved direct numerical simulation (PR--DNS) of fixed particle assemblies is utilized to develop a correlation. The stochastic EL framework specifies unresolved drag force statistics, leading to the correct evolution and sustainment of particle velocity variance over a wide range of Reynolds numbers and solids volume fractions when compared to PR--DNS of freely-evolving homogeneous suspensions. By contrast, standard EL infers drag statistics from variations in the resolved flow and thus under-predicts the growth and steady particle velocity variance in homogeneous suspensions. Velocity statistics from standard EL approaches are found to depend on the bandwidth of the projection function used for two-way momentum coupling, while results obtained from the stochastic EL approach are insensitive to the projection bandwidth. 
\end{abstract}

\maketitle

\section{Introduction}
Eulerian--Lagrangian (EL) methods, in which particles are tracked individually and the fluid is solved on an Eulerian grid, have gained considerable traction for modeling particle-laden flows due to a balance between speed and resolution \citep{cundall_discrete_1979,tsuji_discrete_1993,van_der_hoef_numerical_2008,capecelatro_eulerlagrange_2013}. In recent years, emphasis has been placed on moderate to high mass loading where particles have a first-order effect on the underlying fluid flow disturbances, which feed back to the particle dynamics. Since EL methods do not resolve the fluid boundary layer at the surface of each particle, they enable grid spacings on the order of, or larger than, the particle diameter. The reduced resolution in EL methods requires a model for the hydrodynamic fluid-particle force. By contrast, particle-resolved direct numerical simulation (PR--DNS) resolves the fluid boundary layers around each particle, and thus, the interphase drag force is an output from such simulations. For strongly-coupled flows with inertial particles, increasing the quantitative agreement between EL methods and PR--DNS requires critical assessment of the drag force model. Generally speaking, existing models for drag only capture low-order statistics, such as the mean hydrodynamic force exerted on a particle-laden suspension. It is now recognized that suspensions will exhibit significant variance in the drag force due to interactions between particles and fluid disturbances generated by their neighbors. Variance in drag force, arising from neighbor-induced fluid velocity fluctuations (pseudo-turbulent kinetic energy; PTKE), is generally ignored in EL frameworks. However, recent works have highlighted the importance of PTKE in particle-laden flows; see the closed-form model \citep{mehrabadi_pseudo-turbulent_2015} and transport equations \citep{shallcross_volume-filtered_2020}. 

It has become increasingly well established that standard EL methods, which employ a mean drag closure and neglect PTKE-induced drag disturbances, under-predict drag variance when compared to PR--DNS \cite{kriebitzsch_fully_2013,tenneti_stochastic_2016,akiki_pairwise-interaction_2017}. To this end, multiple PR--DNS studies have demonstrated that flow past a collection of monodisperse spheres yields normally distributed drag forces \citep{akiki_force_2016,esteghamatian_micromeso_2017,huang_effects_2017,subramaniam_towards_2018} whose standard deviation is comparable in magnitude to its mean \cite{akiki_force_2016,huang_effects_2017}. Due to the formation of fluid wakes (PTKE), particles will interact with each other indirectly over length scales comparable to their diameter. These particle-wake interactions give rise to short-range drag perturbations that drive relative motion between neighboring particles, such as drafting-kissing-tumbling (DKT)~\citep{fortes_nonlinear_1987}. Therefore, neglecting higher-order drag statistics, resulting from neighbor-effects, can detrimentally impact EL predictions for higher-order particle statistics (velocity variance and dispersion)~\citep{akiki_pairwise-interaction_2017}. To-date, few drag models have been proposed that account for neighbor-induced disturbances; which may be broadly grouped into deterministic \citep{akiki_pairwise_2017,seyed-ahmadi_microstructure-informed_2020,akiki_shear-induced_2020} and stochastic approaches \citep{tenneti_stochastic_2016,esteghamatian_stochastic_2018,lattanzi_stochastic_2020}. In the former, the drag force experienced by a given particle is directly mapped to its pairwise neighbor interactions, requiring that the relative position of each particle be known when computing the drag force. It is worth noting that the aforementioned information is available in an EL framework but not in an Euler--Euler (EE) framework where the solids are treated as a continuum. By contrast, stochastic approaches aim to capture higher-order particle statistics, resulting from drag fluctuations, without detailed knowledge of each particle position.

Here, a statistical approach is employed to account for neighbor-induced drag fluctuations. Specifically, we follow the mathematical theory derived by \citet{lattanzi_stochastic_2020} for a hierarchy of Langevin equations and employ a stochastic drag force treatment. The interested reader is referred to the cited work for a more inclusive discussion on what the force Langevin (FL) framework yields in terms of particle-phase moments. However, we note that the fluctuating drag statistics obtained from an Ornstein-Uhlenbeck (OU) process were shown to be consistent with drag statistics extracted from freely-evolving PR--DNS of homogeneous systems. Namely, the fluctuating drag is normally distributed with an exponential Lagrangian autocovariance function. As a result of capturing the drag fluctuation statistics, the steady particle velocity variance obtained with FL also agreed with PR--DNS of homogeneously distributed inertial particles at moderate Reynolds numbers. In this manner, the stochastic drag force is designed to reproduce statistics obtained from fully-resolved simulations and is not an empirically added fluctuation.

We emphasize that the present study is centered around a stochastic description of neighbor-induced drag disturbances, where the physical mechanism for drag perturbations is attributed to fluid flow disturbances generated by particles that are in close proximity. It should be noted that the concept of drag force disturbances is not intrinsic to dense suspensions but will also be present in under-resolved simulations of turbulent flows. Specifically, unresolved fluid turbulence also provides a source for particle drag disturbances in dilute flows where neighbor-induced effects are less significant. Pioneering works on turbulent single-phase flows \citep{haworth_generalized_1986,sawford_reynolds_1991,pope_lagrangian_1994,pope_stochastic_2002} have developed Langevin equations for reconstructing the total fluid velocity that may be applied to dilute multiphase flows \citep{iliopoulos_stochastic_2003,pozorski_filtered_2009,pai_two-way_2012}. Therefore, the fluid velocity Langevin model is akin to the force Langevin framework noted above with a crucial difference being that closures for the statistical process are appropriate for dilute turbulent flows without significant two-way coupling. Drag fluctuations arise in the former case from under-resolving the intrinsic fluid turbulence; while in the latter case, they arise from under-resolving the pseudo-turbulence generated by boundary layers around neighboring particles. 
    
The remainder of the paper is arranged as follows. In Sec.~\ref{sec:Fdstats}, a statistical description is introduced for the hydrodynamic force felt by a particle in a dynamical suspension. The drag force is decomposed into a mean and fluctuating component, where the fluctuating component is a stochastic variable that specifies higher-order statistics. In Sec.~\ref{sec:FLframe}, closure is proposed for the fluctuating drag within homogeneous suspensions of inertial particles at moderate Reynolds numbers. Details regarding the numerics are provided in Sec.~\ref{sec:framework}. Homogeneous fluidization of elastic particles is considered in Sec.~\ref{sec:verif} as a canonical flow for comparison to the PR--DNS data of \citet{tenneti_stochastic_2016} and \citet{tavanashad_effect_2019}. Specifically, depending upon the initial conditions, particle velocity variance in homogeneous fluidization will grow (heat) or decay (cool) to a steady state where neighbor-induced drag disturbances are balanced by hydrodynamic dissipation. Therefore, the ability of an EL method to capture the evolution of particle velocity variance in homogeneous fluidization is a significant test of the method's ability to capture higher-order drag statistics. We first demonstrate that the new stochastic EL framework yields convergent velocity variance in homogeneous fluidization, while the velocity variance resulting from standard EL inherently depends upon the length scale employed during two-way coupling (i.e., grid spacing or filter size). Next, the proposed stochastic EL framework is directly compared to PR--DNS data in the fluidized homogeneous heating system (FHHS) and fluidized homogeneous cooling system (FHCS). Finally, we assess the role of neighbor-induced drag fluctuations in a large-scale simulation of cluster-induced turbulence (CIT) in Sec.~\ref{sec:largescale}. 

\section{A statistical description of drag} \label{sec:Fdstats}
Particle motion follows Newton's second law where acceleration results from the net force acting upon the body. When considering the hydrodynamic drag force exerted on particle `$i$,' $\bm{F}_{\rm{inter}}^{(i)}$, an exact integration of the fluid stress tensor $\check{\bm{\tau}}$ over the surface of a particle $\Gamma$ may be evaluated as
\begin{equation}
\bm{F}_{\rm{inter}}^{(i)} =  \int_{\Gamma} \check{\bm{\tau}} \cdot \bm{n} \hspace{0.5ex}{\rm d}A, \label{eq:stressint}
\end{equation}
where $\check{(\cdot)}$ denotes a microscale quantity prior to any averaging, ${\rm d}A$ is an infinitesimal area element, and $\bm{n}$ is the unit normal vector outward from the particle surface. For an \emph{isolated} sphere subject to non-uniform Stokes flow, \citet{maxey_equation_1983} consider a rigorous evaluation of Eq.~\eqref{eq:stressint} that yields contributions from the undisturbed fluid flow, quasi-steady drag, added mass, and Basset history: $\bm{F}_{\rm{inter}}^{(i)} = \bm{F}_{\rm{un}}^{(i)} + \bm{F}_{\rm{qs}}^{(i)} + \bm{F}_{\rm{am}}^{(i)} + \bm{F}_{\rm{uv}}^{(i)}$. However, for a general dynamic suspension, finite Reynolds number effects and neighbor disturbances do not allow a first-principles solution to be obtained. Consequently, standard EL methods generally rely upon drag correlations.

\begin{figure}
    \centering
	\includegraphics[width=0.75\textwidth]{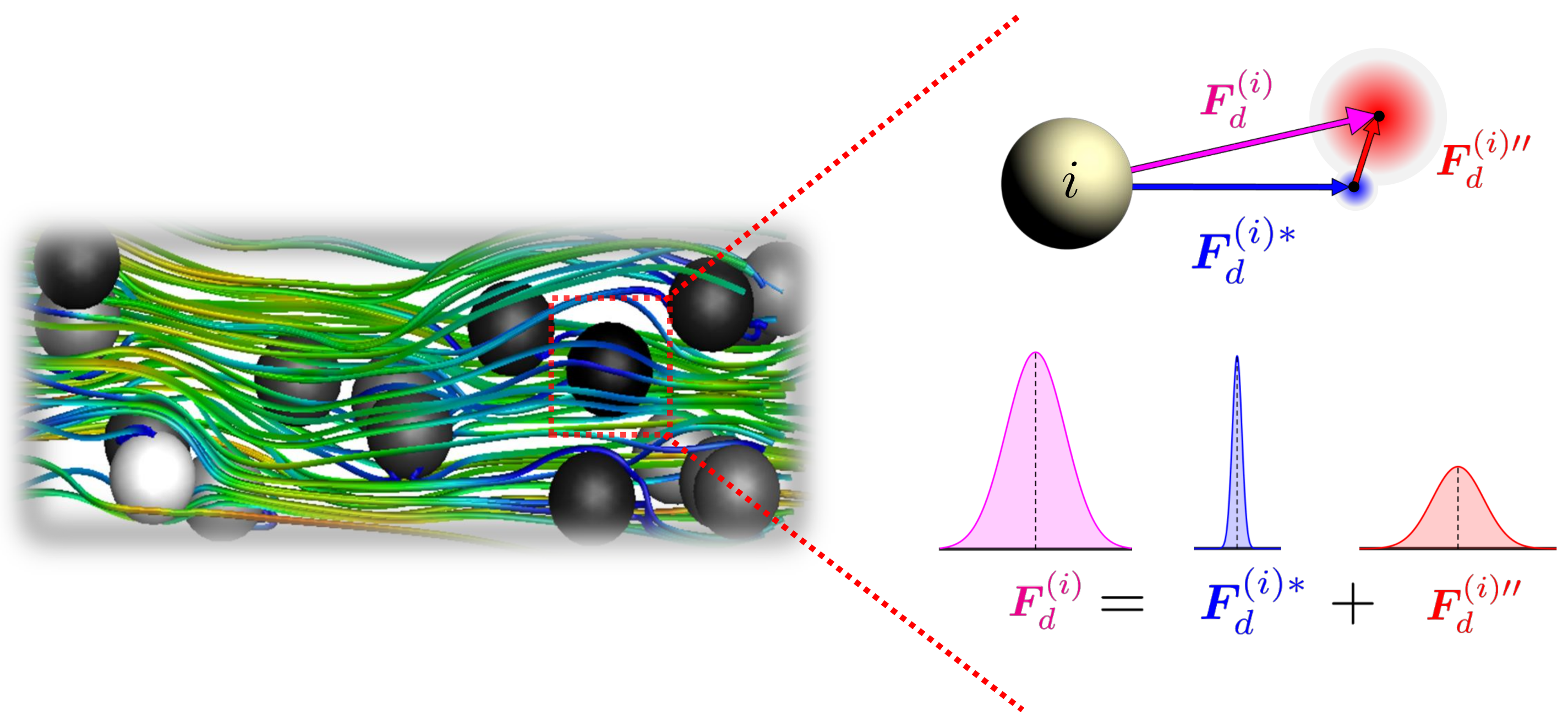}
	\caption{\small{Neighboring particles disturb the local flowfield within a particle-laden suspension, leading to higher-order drag statistics. Negative fluid velocity fluctuations (blue streamlines) correspond to reduced drag forces (darker particles) while positive fluid velocity fluctuations (red streamlines) correspond to higher drag forces (lighter particles). A statistical description is adopted for the drag on an individual particle according to Eq.~\eqref{eq:fdecomp}, where $\bm{F}_{d}^{(i)\prime\prime}$ allows specification of higher-order statistics.}}
\label{fig:subgrid}
\end{figure}

Following \citet{anderson_fluid_1967}, the fluid stress tensor in Eq.~\eqref{eq:stressint} may be decomposed into a filtered component $\bm{\tau}$ and residual $\bm{{\tau}}^{\prime}$, such that $\check{\bm{{\tau}}}=\bm{\tau}+\bm{{\tau}}^{\prime}$. Employing the divergence theorem and choosing a filter length scale such that $\bm{\tau}$ varies little over the volume of the particle, one obtains
\begin{equation}
\int_{\Gamma} \check{\bm{\tau}} \cdot \bm{n} \hspace{0.5ex}{\rm d}A =  \mathcal{V}_{p}^{(i)} \nabla \cdot \bm{\tau} \left[ {\bm{X}_{p}^{(i)}} \right] + \int_{\Gamma} \bm{\tau}^{\prime} \cdot \bm{n} \hspace{0.5ex}{\rm d}A \label{eq:fres},
\end{equation}
where $\mathcal{V}_{p}^{(i)}$ is the volume of the particle and $\bm{\tau} \left[ {\bm{X}_{p}^{(i)}} \right]$ is the filtered stress tensor evaluated at the position of the particle $\bm{X}_{p}^{(i)}$. Traditionally, the unresolved drag force $\int \bm{\tau}^{\prime} \cdot \bm{n} \hspace{0.5ex}{\rm d}A$ is closed with correlations obtained from  experiments \citep{ergun_fluid_1949,wen_mechanics_1966,gidaspow_multiphase_1994} or PR--DNS \citep{hill_moderate-reynolds-number_2001,beetstra_drag_2007,tenneti_drag_2011,rubinstein_lattice_2016}. However, these correlations only capture the average hydrodynamic force acting on a suspension. Consequently, particles that experience the same filtered hydrodynamic environment will experience the same modeled drag force, even though the neighbor-induced flow may cause significant departure from the mean contribution to drag. It should be noted that the first term on the right-hand side of Eq.~\eqref{eq:fres} takes the same form as $\bm{F}_{\rm{un}}^{(i)}$ in the classical formulation of \citet{maxey_equation_1983}, but inherently contains the effects of particle disturbances due to two-way coupling. The second term involving $\bm{\tau}'$ contains all of the other contributions (quasi-steady drag, added mass, and Basset history). Rather than attempting to tease out how each pair-wise neighbor interaction contributes to the hydrodynamic forces on a given particle, we seek a statistical representation that treats these effects stochastically. To build upon the standard EL approach and account for neighbor effects, we expand the unresolved drag into a quasi-steady contribution $\bm{F}_{d}^{(i)*}$ and fluctuating $\bm{F}_{d}^{(i)\prime\prime}$ component according to
\begin{equation}
\int_{\Gamma} \bm{\tau}^{\prime} \cdot \bm{n} \hspace{0.5ex}{\rm d}A =\bm{F}_{d}^{(i)*} +  \bm{F}_{d}^{(i)\prime\prime} \label{eq:fdecomp},
\end{equation}
where the double-prime notation is used here to denote a fluctuation around a modeled term while the single-prime denotes a fluctuation with respect to a filtered quantity. Generally speaking, $\bm{F}_{d}^{(i)\prime\prime}$ corresponds to a perturbation from the quasi-steady drag force. Here, we attribute force perturbations to fluid flow disturbances created by neighboring particles. Therefore, $\bm{F}_{d}^{(i)\prime\prime}$ is a stochastic variable whose statistics, such as distribution and time correlation, are designed to be consistent with PR--DNS of freely-evolving particles. In this manner, $\bm{F}_{d}^{(i)\prime\prime}$ allows higher-order drag statistics, originating from neighbor effects, to be directly enforced within a fluidized suspension (see Fig.~\ref{fig:subgrid}). Closure of the stochastic process utilized to model $\bm{F}_{d}^{(i)\prime\prime}$ is provided in Sec.~\ref{sec:FLframe}. 

When applying the decomposition in Eq.~\eqref{eq:fdecomp} to an EL method, it must be noted that there are differences in how PR--DNS studies and EL methods characterize a system. Specifically, EL methods generally employ instantaneous particle velocities and locally interpolated fluid quantities when evaluating drag force correlations; whereas PR--DNS correlations are derived from ensemble-averaged quantities. To this end, we immediately define the mean Reynolds number employed by PR--DNS
\begin{equation}
{\rm Re}_m = (1-\left\langle \phi \right\rangle) \frac{\rho_f d_p \left\lVert \left\langle \bm{W} \right\rangle\right\rVert}{\mu_f}, \label{eq:rem}
\end{equation}
and the particle Reynolds number employed by EL
\begin{align}
{\rm Re}_p &= (1-\phi) \frac{\rho_f d_p \left\lVert \bm{u}_f \left[ {\bm{X}_{p}^{(i)}} \right] - \bm{U}_{p}^{(i)} \right\rVert}{\mu_f}, \label{eq:rep}
\end{align}
where $\phi$ is the solids volume fraction, $\rho_f$ is the fluid density, $d_p$ is the particle diameter, $\mu_f$ is the dynamic viscosity, $\left\langle \bm{W} \right\rangle = \left\langle \bm{u}_f \right\rangle - \left\langle \bm{U}_{p} \right\rangle$ is the mean slip velocity, and $\bm{u}_f$ and $\bm{U}_p$ are the fluid and particle velocity, respectively. The $\left\langle  \cdot \right\rangle$ notation denotes an ensemble-average while the $\left[ {\bm{X}_{p}^{(i)}} \right]$ notation is suppressed on the solids volume fraction for readability. Therefore, the $\bm{F}_{d}^{(i)*}$ contribution in EL, computed with the particle Reynolds number, will be an approximation to the ensemble-averaged mean drag force exerted on a suspension $\left\langle \bm{F}_{d} \right\rangle$. More precisely, $\bm{F}_{d}^{(i)*}$ is a stochastic model for the mean hydrodynamic force, arising from one-particle statistics (position and velocity) in a filtered realization of the fluid field, that is based on a form of average drag from PR--DNS. Therefore, $\bm{F}_{d}^{(i)*}$ is inherently coupled to the momentum filtering employed by EL and will approach a delta function centered at $\left\langle \bm{F}_{d} \right\rangle$ in the limit of infinite filter width; but for finite filter length scales, $\bm{F}_{d}^{(i)*}$ will have finite variance due to variation of interpolated quantities and particle velocity (see Sec.~\ref{subsec:Con}). In the present work, focus is given to filter widths larger than a particle diameter $\delta_f = \mathcal{O}( 10 d_p)$ that do not contribute to significant variation of $\bm{F}_{d}^{(i)*}$ in homogeneous fluidization. To avoid confusion when discussing mean drag, we draw an analogy with the Maxey-Riley equation and refer to $\bm{F}_{d}^{(i)*}$ as the quasi-steady drag force. We further motivate this analogy by noting that the present work focuses on inertial particle suspensions where added mass and Basset history effects are less significant and PR--DNS correlations are obtained from fixed particle simulations. 

\section{Force Langevin framework} \label{sec:FLframe}
The force Langevin (FL) theory examined by \citet{lattanzi_stochastic_2020} is employed here to describe the neighbor-induced fluctuating drag force $\bm{F}_{d}^{(i)\prime\prime}$. Specifically, $\bm{F}_{d}^{(i)\prime\prime}$ follows an Ornstein-Uhlenbeck (OU) process according to
\begin{equation}
\mathrm{d}\bm{F}_{d}^{(i)\prime\prime} =  -\frac{1}{\tau_{F}} \bm{F}_{d}^{(i)\prime\prime} \mathrm{d}t + \frac{\sqrt{2}\sigma_F}{\sqrt{\tau_{F}}} \mathrm{d}\bm{W}_t, \label{eq:pfl}
\end{equation}
where $\tau_F$ is the integral time scale of the fluctuating drag force, $\sigma_F$ is the standard deviation of the fluctuating drag force, and $\mathrm{d}\bm{W}_t$ is a Wiener process increment. We refer the interested reader to \citet{lattanzi_stochastic_2020} where a detailed discussion regarding motivation for, and results from, the OU process are provided. Here, we briefly emphasize that the steady solution to the OU process is a normal distribution $\mathcal{N}\left[0, \hspace{0.5ex} \sigma_{F}\right]$ and a multitude of PR--DNS studies have reported normally distributed drag forces $\mathcal{N}\left[\left\langle F_{d}\right\rangle, \hspace{0.5ex} \sigma_{F}\right]$ \citep{kriebitzsch_fully_2013,akiki_force_2016,huang_effects_2017,esteghamatian_stochastic_2018,subramaniam_towards_2018}. Additionally, FL was shown to be reconcilable with PR--DNS results for granular temperature evolution since it correctly attributes the source of granular temperature to the force-velocity covariance \citep{lattanzi_stochastic_2020}.

In general, coefficients of the OU process in Eq.~\eqref{eq:pfl}, namely the drag time scale $\tau_F$ and standard deviation $\sigma_F$, may be tensorial quantities that correlate the drag fluctuations in each direction and drive anisotropic granular temperature development. While a valuable and interesting problem, we first consider the isotropic case given in Eq.~\eqref{eq:pfl} so that the same force time scale and standard deviation is applied to all three fluctuating force directions. We further note that particle collisions will relax the granular temperature back towards isotropy. For statistically homogeneous suspensions, \citet{garzo_enskog_2012} derived an evolution equation for the particle velocity anisotropy tensor that shows a return to isotropy occurs at steady state due to particle collisions. For dilute particle flows with small density ratio $\rho_p/\rho_f = \mathcal{O}(1)$, where viscous lubrication is significant and particle collisions are less frequent, anisotropy may play a significant role. However, a description for such flows will need to be addressed in future work as the focus here is on inertial gas-solid suspensions.

\subsection{Force time scale} \label{subsec:tauf}
In principle, the fluctuating force time scale $\tau_{F,ij}$ characterizes the memory of temporal correlation of the drag force fluctuation $\bm{F}_{d}^{\prime\prime}$ and may be extracted from PR--DNS simulation with freely-evolving particles via the Lagrangian covariance function
\begin{equation}
\tau_{F,ij} = \int_0^{\infty} \frac{\left\langle F_{d,i}^{\prime\prime}(t+s)F_{d,j}^{\prime\prime}(t)\right\rangle}{\left\langle F_{d,i}^{\prime\prime}(t)F_{d,j}^{\prime\prime}(t)\right\rangle} \hspace{0.5ex} \mathrm{d}s, \label{eq:kff}
\end{equation}
where we note there is no sum over repeated force indices within the bracketed terms. The analysis by \citet{lattanzi_stochastic_2020} shows that FL generates a fluid-mediated source to particle velocity fluctuations through the force-velocity covariance. Specifically, the fluctuating drag forces drive the development of particle velocity fluctuations, and the time over which the fluctuating drag and particle velocity become correlated will dictate the magnitude of granular temperature that can be developed. 

In a dynamic suspension, particles will experience states of free-streaming and redirection due to collisions. During free-streaming, particles accelerate in the direction of the fluctuating drag force, developing force-velocity correlation \citep{lattanzi_stochastic_2020}. However, interparticle collisions will act to redirect the particle velocity vector and decorrelate the fluctuating force and velocity. For inertial particles, the mean-free time between collisions, $\tau_{\rm col}$, is expected to be a good approximation to the force-velocity correlation time scale. Agreement between $\tau_F$ and $\tau_{\rm col}$ was demonstrated in \citet{lattanzi_stochastic_2020}. It is noteworthy that the deterministic Euler-Euler model developed by \citet{koch_particle_1999} for high Stokes number particles also considers the mean-free time as an approximation for $\tau_F$, but is restricted to low Reynolds number regimes. From these physical arguments, and previous numerical results, we approximate the time scale for the fluctuating hydrodynamic force with the mean-free time between successive collisions \citep[see ch. 5 of][]{chapman_mathematical_1970}
\begin{equation}
\tau_F \approx \tau_{\rm col} = \frac{d_p}{24 \phi \chi} \sqrt{\frac{\pi}{\Theta}},
\label{eq:mft}
\end{equation}
where $\Theta$ is the granular temperature and $\chi$ is the radial distribution function at contact, given by~\citep{MA1988191}
\begin{equation}
\chi = \frac{1 + 2.50\phi + 4.51\phi^2 + 4.52\phi^3}{\left[1 - \left(\phi/0.64\right)^3\right]^{0.68} }.
\label{eq:rdf}
\end{equation}
At this point, two things should be noted. First, $\tau_{\rm col}$ is derived from the kinetic theory of non-uniform gases and thus is agnostic to PR--DNS data. 
Second, the granular temperature $\Theta$ corresponds to spatially \emph{uncorrelated} velocity fluctuations while the particle velocity variance $T$ does not employ spatial conditioning \cite{fevrier_partitioning_2005}. These two quantities are equivalent for homogeneous systems, where the spatially correlated mean equals the domain average, but not in inhomogeneous systems. We distinguish the granular temperature and particle velocity variance in nomenclature and mathematical computation; see Secs.~\ref{subsec:gtcomp} and \ref{sec:verif} for more details regarding $\Theta$ and $T$, respectively.

\subsection{Force standard deviation}  \label{subsec:sigf}
In this section, the standard deviation in drag force is evaluated from PR--DNS of homogeneous fluid-particle suspensions. It has already been well established that the mean drag force exerted on a suspension of high inertia particles ($\rho_p/\rho_f \geq \mathcal{O}(100)$) is well approximated by PR--DNS simulation of static particle assemblies \citep{hill_moderate-reynolds-number_2001,beetstra_drag_2007,tenneti_drag_2011}. Moreover, \citet{tavana_ijmf_2021} recently performed PR--DNS with fixed and freely-evolving particles over a wide range of conditions, including density ratio and Reynolds number, to study the effect of particle mobility on drag. It was shown that the corresponding drag correlation converges to the fixed bed correlation of \citet{tenneti_drag_2011} for large density ratios. In a similar fashion, a correlation for the standard deviation in drag force from the PR--DNS dataset of \citet{tavana_ijmf_2021} is pursued here.  As was seen with mean drag, we observe convergence of the standard deviation in drag force $\sigma_F$, extracted from PR--DNS with freely-evolving and fixed particle suspensions, at large particle-to-fluid density ratios $\rho_p/\rho_f \geq 100$. For this reason, a correlation is obtained from simulations with \emph{fixed} particle assemblies. In the dataset of \citet{tavana_ijmf_2021}, a simulation is defined by the mean solids volume fraction and mean Reynolds number. For a given set of conditions $\left({\rm Re}_m,\, \left\langle \phi \right\rangle \right)$, five particle configurations were generated, each containing 200 particles. In each realization, the mean drag force and standard deviation in drag force were computed. Finally, the force standard deviation was ensemble-averaged over all realizations.

\begin{figure}
    \centering
	\includegraphics[width=0.35\textwidth]{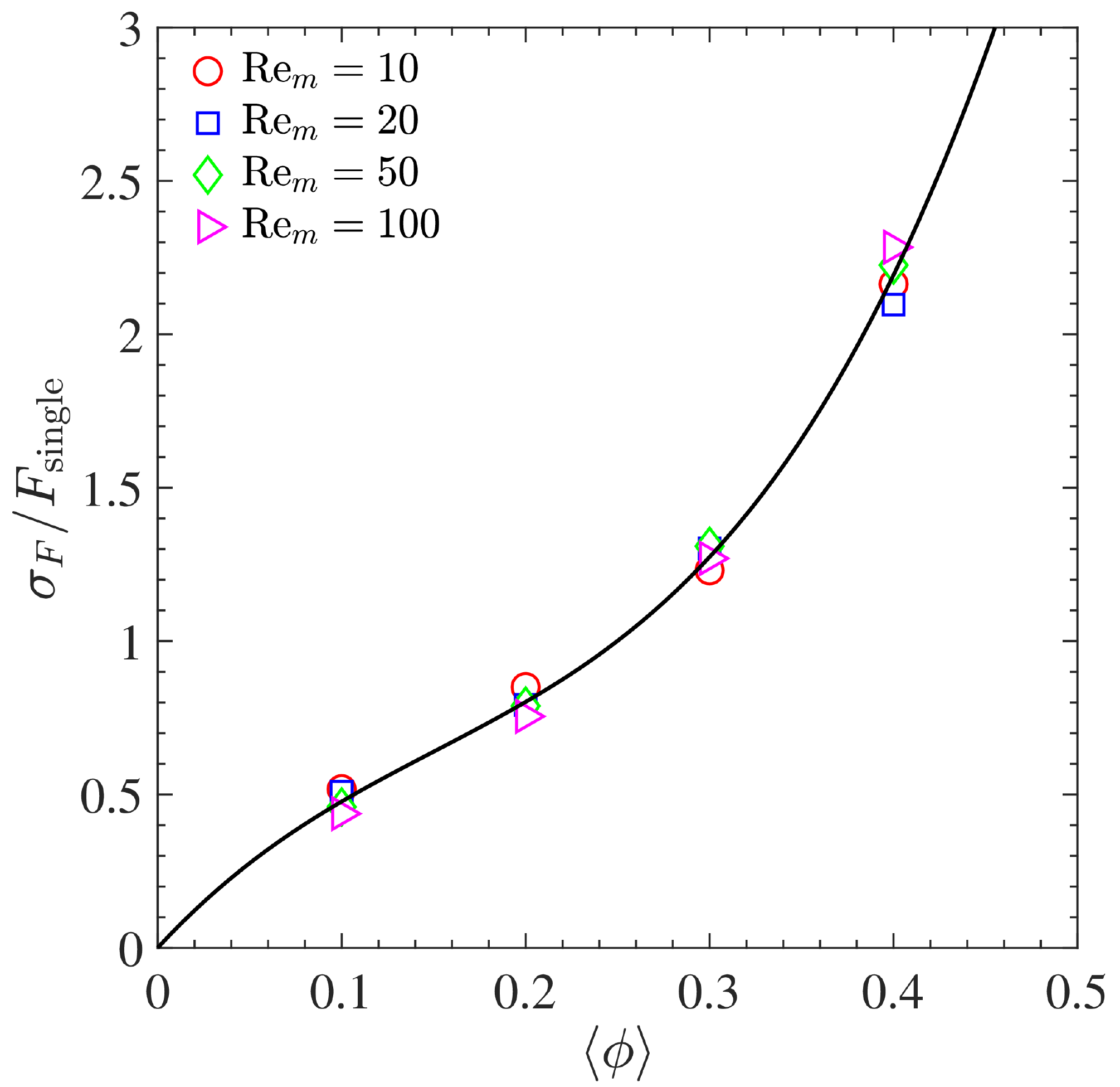}
	\caption{\small Collapse of the drag force standard deviation $\sigma_F$ obtained from PR--DNS when normalized by the drag force on an isolated sphere $F_{\rm single}$. The black line corresponds to the correlation given by Eq.~\eqref{eq:fvarcor}.}
\label{fig:sigF}
\end{figure}

When developing a mean drag force correlation for inertial particles, \citet{tenneti_drag_2011} showed that a solids volume fraction correction could be obtained for the normalized drag force $\left\langle  \bm{F}_d \right\rangle/F_{\mathrm{single}}$, where 
\begin{align}
F_{\mathrm{single}} &= 3\pi \mu_f d_p \left( 1 + 0.15 {\rm Re}_m^{0.687} \right) \left(1- \left\langle \phi \right\rangle \right) \left\lVert \left\langle \bm{W} \right\rangle \right\rVert \label{eq:Fsingle}
\end{align}
is the drag force on an isolated sphere given by the classic Schiller and Naumann correlation~\cite{clift_bubbles_2013} and evaluated using the mean slip velocity. In a similar fashion, collapse of the normalized standard deviation $\sigma_F/F_{\mathrm{single}}$ is observed here, albeit onto a different function of solids volume fraction $f_{\phi}^{\sigma_F}$ (see Fig.~\ref{fig:sigF}). A third-order polynomial in solids volume fraction is fit to the data to obtain
\begin{align}
f_{\phi}^{\sigma_F} &= 6.52\phi - 22.56 \phi^2 + 49.90 \phi^3, \label{eq:fvarcor}
\end{align}
which ensures that $\lim_{\phi \rightarrow 0} \sigma_F = 0$, i.e., there will be no neighbor-induced drag perturbations in the infinitely dilute limit. Thus, the standard deviation in drag may be readily modeled in EL as
\begin{subequations}
\begin{align}
\frac{\sigma_F}{m_{p}^{(i)}} \equiv f_{\phi}^{\sigma_F} \frac{F_{\mathrm{single}}}{m_{p}^{(i)}} &= f_{\phi}^{\sigma_F} f_{\rm iso} \frac{(1-\phi)\left\lVert \bm{u}_f \left[ {\bm{X}_{p}^{(i)}} \right] - \bm{U}_{p}^{(i)}\right\rVert}{\tau_p}, 
\label{eq:fvarcomp} \\[1.0ex]
f_{\rm iso} &= \left( 1 + 0.15 {\rm Re}_p^{0.687} \right)
\end{align}
\end{subequations}
where $m_p^{(i)}=\rho_p\mathcal{V}_p^{(i)}$ is the particle mass and $\tau_p = \rho_p d_p^2/(18 \mu_f)$ is the Stokes response time. 

Examination of Eqs.~\eqref{eq:fvarcor}-\eqref{eq:fvarcomp} shows that the force standard deviation is written in terms of locally filtered fields and instantaneous particle velocities, $\phi; \,{\rm Re}_p$, so as to make it applicable to EL. Specifically, the correlation is developed from PR--DNS data with ensemble-averaged quantities $\left\langle \cdot \right\rangle$ but is intended for use in EL, see discussion at the end of Sec.~\ref{sec:Fdstats}. In the absence of a more formal route for extending PR--DNS derived correlations to EL methods, we proceed in the standard manner by replacing ensemble-averaged quantities with their locally filtered counterpart.

\section{Euler--Lagrange framework}  \label{sec:framework}
In this section, the stochastic EL framework and its discretization are summarized, using closures reported in Sec.~\ref{sec:FLframe} to model the drag force perturbations.

\subsection{Particle-phase description} \label{subsec:part}
The translational motion of each particle follows Newton's second law according to
\begin{subequations}
\begin{align}
\frac{\mathrm{d}\bm{X}_{p}^{(i)}}{\mathrm{d}t} &= \bm{U}_{p}^{(i)}, \label{eq:ppos}\\[1.0ex]
m_{p}^{(i)}\frac{\mathrm{d}\bm{U}_{p}^{(i)}}{\mathrm{d}t} &= \sum_{j=1}^{N} \bm{F}_{{\rm col}}^{(ij)} + \bm{F}_{\rm{inter}}^{(i)} + m_{p}^{(i)}\bm{g}. \label{eq:pvel}
\end{align}
\end{subequations}
A soft-sphere approach is employed for the collisional force $\bm{F}_{\rm col}$, where each particle contact is described as a linear-spring-dashpot~\citep{cundall_discrete_1979}. The simulation timestep $\Delta t$ is restricted such that particles do not move more than one-tenth of their diameter per timestep, thereby avoiding excessive overlap \citep{capecelatro_eulerlagrange_2013}. A second-order Runge--Kutta (RK) scheme is utilized to integrate Eqs.~\eqref{eq:ppos}--\eqref{eq:pvel} in time. Combining Eqs.~\eqref{eq:stressint}--\eqref{eq:fdecomp}, the interphase momentum exchange term $\bm{F}_{\rm{inter}}^{(i)}$ contains contributions from the resolved stress, quasi-steady drag, and fluctuating drag according to
\begin{equation}
\bm{F}_{\rm{inter}}^{(i)} = \mathcal{V}_{p}^{(i)} \nabla \cdot \bm{\tau} \left[ {\bm{X}_{p}^{(i)}} \right] + \bm{F}_{d}^{(i)*} +  \bm{F}_{d}^{(i)\prime\prime}. \label{eq:finter}
\end{equation}
Since we are focusing on inertial particle suspensions, the quasi-steady drag closure provided in \citet{tenneti_drag_2011} is employed here
 \begin{equation}
\frac{\bm{F}_{d}^{(i)*}}{m_{p}^{(i)}} = \left( \frac{f_{\mathrm{iso}}}{(1-\phi)^2} + f_{\phi} + f_{\phi,{\rm Re}_p}  \right) \frac{(1-\phi)\left(\bm{u}_f \left[ {\bm{X}_{p}^{(i)}} \right] - \bm{U}_{p}^{(i)}\right)}{\tau_p},
\label{eq:tenn}
\end{equation}
where
\begin{subequations}
\begin{align}
f_{\phi} &= \frac{5.81 \phi}{(1-\phi)^2} + 0.48\frac{\phi^{1/3}}{(1-\phi)^3}, \\[1.0ex]
f_{\phi,{\rm Re}_p} &= \phi^3 (1-\phi) {\rm Re}_p \left( 0.95 + \frac{0.61 \phi^3}{(1-\phi)^2}\right). \label{eq:tenn3}
\end{align}
\end{subequations}

The fluctuating drag force is updated in time according to Eq.~\eqref{eq:pfl}. However, special care needs to be taken when integrating stochastic differential equations, as classical schemes for deteriministic ordinary differentials are not strongly consistent for stochastic differentials \cite{kloeden_numerical_1992}. Integration of Eq.~\eqref{eq:pfl} is handled via an explicit RK scheme that exhibits first-order strong and weak convergence \cite[see Ch. 11 of][]{kloeden_numerical_1992}
\begin{align}
F_{d,k}^{\prime\prime \, n+1} &= \left(1 - a^{n}\right) F_{d,k}^{\prime\prime \, n} \Delta t + b^{n} \Delta W_t + \frac{\Delta W_t^2- \Delta t}{2\sqrt{\Delta t}} \left[ b^{n+1/2} - b^{n} \right], \label{eq:fldisc}
\end{align}
where $F_{d,k}^{\prime\prime \, n}$ is the $k$-th component of the fluctuating force at the $n$-th time iteration, $a = 1/\tau_F$, $b =  \sqrt{2}\sigma_F/\sqrt{\tau_{F}}$, and $\Delta W_t = \sqrt{\Delta t} \hspace{0.5ex} \mathcal{N}\left[0, \hspace{0.5ex} 1 \right]$. As shown in Sec.~\ref{sec:FLframe}, the coefficients of the OU process in Eq.~\eqref{eq:pfl} are only a function of hydrodynamic variables. Therefore, $b^{n+1/2}$ is evaluated at the midpoint step resulting from second-order RK integration of the particle position and velocity. The third term on the right-hand side of Eq.~\eqref{eq:fldisc} corresponds to a finite difference approximation for spatial variation in $b$ while the other two terms comprise the standard Euler--Maruyama method. Therefore, Eq.~\eqref{eq:fldisc} will simplify to the Euler integration scheme for homogeneous OU coefficients. Initialization of the fluctuating force is achieved by sampling from the steady Fokker-Planck solution $\mathcal{N}\left[0, \hspace{0.5ex} \sigma_{F}\right]$. Since a soft-sphere collision model is employed here, the time step $\Delta t$ required to accurately resolve collisions guarantees the OU stability criterion $\Delta t \le \tau_F$.

\subsection{Fluid-phase description} \label{subsec:fluid}
In order to account for two-way coupling between the fluid and particle phases, without resolving the boundary layers around each particle, we consider a volume-filtered description for the fluid phase. Specifically, the pointwise Navier--Stokes equations are replaced with locally smoothed conservation equations for mass and momentum \citep{anderson_fluid_1967}
\begin{subequations}
\begin{align}
\frac{\partial}{\partial t}\left((1-\phi)\rho_f\right) + \nabla \cdot \left((1-\phi)\rho_f \bm{u}_f \right) &= 0, \label{eq:con}\\[1.0ex]
\frac{\partial}{\partial t}\left((1-\phi)\rho_f\bm{u}_f\right) + \nabla \cdot \left((1-\phi)\rho_f \bm{u}_f \otimes \bm{u}_f \right) &= \nabla \cdot \bm{\tau} + (1-\phi)\rho_f \bm{g} - \bm{\mathcal{F}}_{\rm inter} + \bm{\mathcal{F}}_{\rm mfr}, \label{eq:mom}
\end{align}
\end{subequations}
where $\bm{g}$ is the gravitational body force and $\bm{\mathcal{F}}_{\rm mfr}$ is a forcing term utilized to establish a desired mass flow rate. The fluid stress tensor $\bm{\tau}$ is given by
\begin{equation}
\bm{\tau} = -p_f\bm{I} + \mu_f \left( \nabla \bm{u}_f + \nabla \bm{u}_f^{\intercal} - \frac{2}{3} (\nabla \cdot \bm{u}_f)\bm{I} \right), \label{eq:fstress}
\end{equation}
where $p_f$ is the pressure and $\bm{I}$ is the identity matrix. The source term due to interphase momentum transfer $\bm{\mathcal{F}}_{\rm inter}$ is obtained by projecting each particle's hydrodynamic force onto the fluid mesh and is discussed in detail in Sec.~\ref{subsec:fpcouple}.

The fluid phase conservation equations are solved using NGA \citep{desjardins_high_2008}, a fully conservative, low-Mach number finite volume solver. A staggered grid with second-order spatial accuracy is advanced in time with the semi-implicit Crank--Nicolson scheme of \citet{pierce_progress-variable_2001} while the pressure Poisson equation is solved via fast Fourier transforms to enforce continuity. We emphasize that the particle and fluid equations are staggered in time such that the particle equations are solved using the fluid velocity obtained from an Adams-Bashforth predictor step, which is second-order accurate, resulting in second-order temporal accuracy for the coupling between particles and fluid phase.

\subsection{Two-way coupling} \label{subsec:fpcouple}
The projection of particle data onto the Eulerian grid is performed by way of a Gaussian kernel $\mathcal{G}\left( \left\lVert \bm{x} - \bm{X}_{p}^{(i)} \right\rVert \right)$
\begin{subequations}
\begin{align}
\phi &= \sum_{i=1}^{N_p} \mathcal{G}\left( \left\lVert \bm{x} - \bm{X}_{p}^{(i)} \right\rVert \right)\mathcal{V}_{p}^{(i)}, \label{eq:projphi}\\[1.0ex]
\bm{\mathcal{F}}_{\rm inter} &= \sum_{i=1}^{N_p} \mathcal{G}\left(  \left\lVert \bm{x} - \bm{X}_{p}^{(i)} \right\rVert \right) \bm{F}_{\rm{inter}}^{(i)}, \label{eq:projfinter}
\end{align}
\end{subequations}
with characteristic size $\delta_f$ that corresponds to the full width at half height. Unless otherwise stated, the Gaussian kernel in this work was held fixed at $\delta_f = 7d_p$. The operations given in Eqs.~\eqref{eq:projphi}--\eqref{eq:projfinter} are computed efficiently by a two-step filtering procedure where the particle data is first extrapolated to the nearest grid points on the fluid mesh and then diffused implicitly to the characteristic width of the kernel $\delta_f$ \citep{capecelatro_eulerlagrange_2013}, thereby yielding Eulerian fields that are spatially smooth.

\subsection{Granular temperature computation} \label{subsec:gtcomp}
While the standard deviation in drag force in Eq.~\eqref{eq:fvarcomp} is straighforward to compute with the resolved fields in an EL simulation, the mean-free time in Eq.~\eqref{eq:mft} is more challenging due to its implicit dependence on the granular temperature, $\Theta$. Since the physical mechanism for force-velocity decorrelation is attributed to collisions, the random uncorrelated particle motion must be utilized in Eq.~\eqref{eq:mft} \cite{fevrier_partitioning_2005}. Special care needs to be taken when evaluating the mean particle velocity about which the fluctuation is defined.  \citet{capecelatro_fluidparticle_2015} showed that in flows with significant heterogeneity in volume fraction (i.e., clustered flows), defining the mean particle velocity using the same procedure employed for two-way coupling results in a strong dependence on the choice of filter size $\delta_f$. Instead, an accurate representation of the spatially correlated velocity can be obtained using an adaptive filter that dynamically adjusts its sampling volume such that a sufficient number of particles are evaluated at each spatial location. The adaptive filter is given by~\citep{capecelatro_fluidparticle_2015}
\begin{equation}
\delta_{\Theta}(\phi) = \left(\frac{\mathcal{N}_p d_p^3}{\phi}\right)^{1/3},
\label{eq:filterwidth}
\end{equation}
to compute the granular temperature over $\mathcal{N}_p = 10$ nearest neighbors. For clarity, we note that $\delta_f$ denotes a fixed Gaussian filter width employed for two-way coupling of momentum but $\delta_{\Theta}$ denotes a variable Gaussian filter width for computing the granular temperature. Projecting particle data with the variable Gaussian filter $\mathcal{G}_{\Theta}$ leads to 
\begin{subequations}
\begin{align}
\phi_{\Theta} &= \sum_{i=1}^{N_p} \mathcal{G}_{\Theta}\left(  \left\lVert \bm{x} - \bm{X}_{p}^{(i)} \right\rVert \right)\mathcal{V}_{p}^{(i)}, \label{eq:projphiT}\\[1.0ex]
\phi_{\Theta} \bm{\mathcal{U}}_{p} &= \sum_{i=1}^{N_p} \mathcal{G}_{\Theta}\left(  \left\lVert \bm{x} - \bm{X}_{p}^{(i)} \right\rVert \right) \mathcal{V}_{p}^{(i)} \bm{U}_{p}^{(i)}. \label{eq:projupT}
\end{align}
\end{subequations}
Defining the particle velocity fluctuation $\delta \bm{U}_{p}^{(i)}$ with respect to the local phasic average
\begin{align}
\delta \bm{U}_{p}^{(i)} &= \bm{U}_{p}^{(i)} - \frac{\phi_{\Theta} \bm{\mathcal{U}}_{p}\left[ {\bm{X}_{p}^{(i)}} \right] }{\phi_{\Theta}\left[ {\bm{X}_{p}^{(i)}} \right]}, \label{eq:DelU}
\end{align}
allows the granular temperature at the particle to be computed as
\begin{align}
\Theta\left[ {\bm{X}_{p}^{(i)}} \right] &= \frac{1}{3} \delta \bm{U}_{p}^{(i)}  \cdot \delta \bm{U}_{p}^{(i)}. \label{eq:compT}
\end{align}
In summary, the adaptive filtering utilized to obtain Eq.~\eqref{eq:compT} has been previously shown to accurately replicate two-point Lagrangian statistics (radial distribution function and velocity autocorrelation)\cite{capecelatro_fluidparticle_2015} and allows for the computation of uncorrelated particle motion within inhomogeneous flows, since it relies upon a local average of the particle velocity. For homogeneous flows, the granular temperature is equivalent to the particle velocity variance $T$; see additional discussion in Sec.~\ref{sec:verif}. The same two--step filtering process in Sec.~\ref{subsec:fpcouple} is employed for Eqs.~\eqref{eq:projphiT}--\eqref{eq:projupT}, with the key difference being that the diffusion coefficient is now a spatially varying quantity (see Eq.~\eqref{eq:filterwidth}).

\section{Homogeneous fluidization of elastic particles} \label{sec:verif}
We consider the homogenenous fluidization of particles within a triply periodic cube of length $L$. A mean fluid flow rate $ \left\langle \bm{u}_{f} \right\rangle = \left[ 0 \hspace{1ex} 0 \hspace{1ex} u_{f,z}\right]^{\top}$ is imposed via $\mathcal{F}_{\rm mfr}$ in Eq.~\eqref{eq:mom} to obtain a desired mean Reynolds number ${\rm Re}_m$, while the gravitational body force $\bm{g} = \left[ 0 \hspace{1ex} 0 \hspace{1ex} -g_z\right]^{\top}$ is prescribed in the opposite direction. $\mathcal{F}_{\rm mfr}$ is added as a uniform source term and is computed each timestep to enforce the constant $u_{f,z}$. The body force $g_z$ is set equal to the mean hydrodynamic acceleration $\left\langle\bm{F}_{d}\right\rangle/m_{p}$, computed via Eq.~\eqref{eq:tenn} for the mean conditions ${\rm Re}_m$ and $\left\langle \phi \right\rangle$. For the homogeneous cases considered here, the drag force offsets the weight of the suspension, leading to a mean particle velocity that is approximately zero throughout the simulation. In contrast, the particle velocity variance will grow or decay to a steady value that is dictated by the balance of drag variation and hydrodynamic dissipation. The simulation domain closely reflects the PR--DNS studies of~\citet{tenneti_stochastic_2016} and \citet{tavanashad_effect_2019}, which serve as benchmark data for comparison to the stochastic EL method presented here. Unless otherwise noted, the grid spacing $\Delta x/d_p = 0.5$, kernel width $\delta_f/d_p = 7$, and domain size $L/d_p = 7$ were held fixed for the homogeneous fluidization simulations, and a summary of relevant conditions is provided in Table~\ref{tab:params}.

Within the homogeneous fluidization system, we define two canonical flows that result from different initial conditions for the particle velocity. Namely, we first examine the fluidized homogeneous heating system (FHHS) in Sec.~\ref{subsec:FHHS}, where particles are initialized with zero velocity and particle velocity variance grows to a steady value. We then examine the fluidized homogeneous cooling system (FHCS) in Sec.~\ref{subsec:FHCS}, where particles are initialized with an over-prescribed velocity variance that decays to a steady value. To match the FHCS simulations completed by~\citet{tenneti_stochastic_2016}, we sample the initial particle velocities from a Maxwellian distribution.

It is important to note that domain sizes $L$ considered by the PR--DNS studies of \citet{tenneti_stochastic_2016} and \citet{tavanashad_effect_2019} are sufficiently small that particle clustering is not observed, and the system remains homogeneous. It has been well established that large-scale fluidized systems exhibit the classic clustering instability due to two-way momentum coupling and/or dissipative collisions. We first focus on the homogeneous case and define crucial parameters to facilitate comparison between EL and PR--DNS benchmark data. Specifically, particle velocity variance $T$ is computed with a velocity fluctuation 
\begin{align}
\bm{U}_{p}^{(i) \prime} &= \bm{U}_{p}^{(i)} -  \left\langle\bm{U}_{p}^{(i)} \right\rangle, \label{eq:DelU2}
\end{align}
that is defined with respect to the domain average $ \left\langle\bm{U}_{p}^{(i)} \right\rangle$, leading to
\begin{align}
T &= \frac{1}{3} \left\langle \bm{U}_{p}^{(i) \prime} \cdot \bm{U}_{p}^{(i) \prime} \right\rangle. \label{eq:Thomo}
\end{align}
Utilizing the definition for particle velocity fluctuation in Eq.~\eqref{eq:DelU2} and particle velocity variance in Eq.~\eqref{eq:Thomo}, we define the particle Reynolds stress tensor $\bm{R}^{p}$, anisotropy tensor $\bm{b}^{p}$, and thermal Reynolds number ${\rm Re}_{T}$ as 
\begin{subequations}
\begin{align}
\bm{R}^{p} &= \left\langle \bm{U}_p^{\prime} \otimes \bm{U}_p^{\prime} \right\rangle, \label{eq:rpp} \\[1.0ex]
\bm{b}^{p} &= \frac{\bm{R}^{p}}{3T} - \frac{1}{3}\bm{{\rm I}},  \label{eq:b} \\[1.0ex]
\end{align}
\end{subequations}
and
\begin{align}
{\rm Re}_{T}      &= \frac{\rho_f d_p \sqrt{T}}{\mu_f}.  \label{eq:ret}
\end{align}
We emphasize that the mean-free time $\tau_{\rm col}$ in Eq.~\eqref{eq:mft} is always computed with the granular temperature $\Theta$ via Eq.~\eqref{eq:compT}, making it valid for inhomogeneous flows. However, when comparing to PR--DNS of homogeneous systems, we report Eqs.~\eqref{eq:rpp}--\eqref{eq:ret} with the particle velocity variance $T$, computed via Eq.~\eqref{eq:Thomo}, for consistency with the definitions employed in those studies. In Sec.~\ref{sec:largescale} we finally consider large-scale simulations of cluster-induced turbulence (CIT) that are characterized by strong inhomogeneities in particle number density and clusters that fall faster than their terminal velocity.

\renewcommand{\arraystretch}{1.3}
\begin{table}
\caption{\small{Simulation conditions}}
\label{tab:params}
\begin{center}
\begin{tabular}{p{1cm} p{2.5cm} }
\hline
\hline
$d_p$   & $500 \times 10^{-6}$ m  \\
$\mu_f$ & $1.0 \times 10^{-5}$ Pa$\cdot$s \\
$\rho_f$ & 1.0 kg/m$^3$ \\
$\left\langle \phi \right\rangle$ & $\left[0.1 \hspace{2ex} 0.4 \right]$ \\
${\rm Re}_m$ & $\left[10 \hspace{2ex} 100 \right]$ \\
$\rho_p/\rho_f$ & $\left[100 \hspace{1ex} 1000 \right]$ \\
\hline
\hline
\end{tabular}
\end{center}
\end{table}

\subsection{Convergence} \label{subsec:Con}
Before drawing detailed comparisons between the new stochastic EL framework, standard EL framework, and PR--DNS, the convergence properties of EL with $\delta_f$ variation are examined within the FHHS. As discussed in Sec.~\ref{subsec:fpcouple}, $\delta_f$ sets the filter length scale for two-way coupling and is therefore directly related to variation in the fluid fields through the projection of $\mathcal{V}_p^{(i)}$ and $\bm{F}_{\rm inter}^{(i)}$. Variation in $\phi$ and $\bm{u}_f$ will lead to variation in the quasi-steady drag contribution $\bm{F}_{d}^{(i)*}$, thereby providing a mechanism for the generation of particle velocity variance.  

\begin{figure}
    \centering
    \begin{subfigure}{0.31\textwidth}
        \centering
        \includegraphics[width=0.99\textwidth]{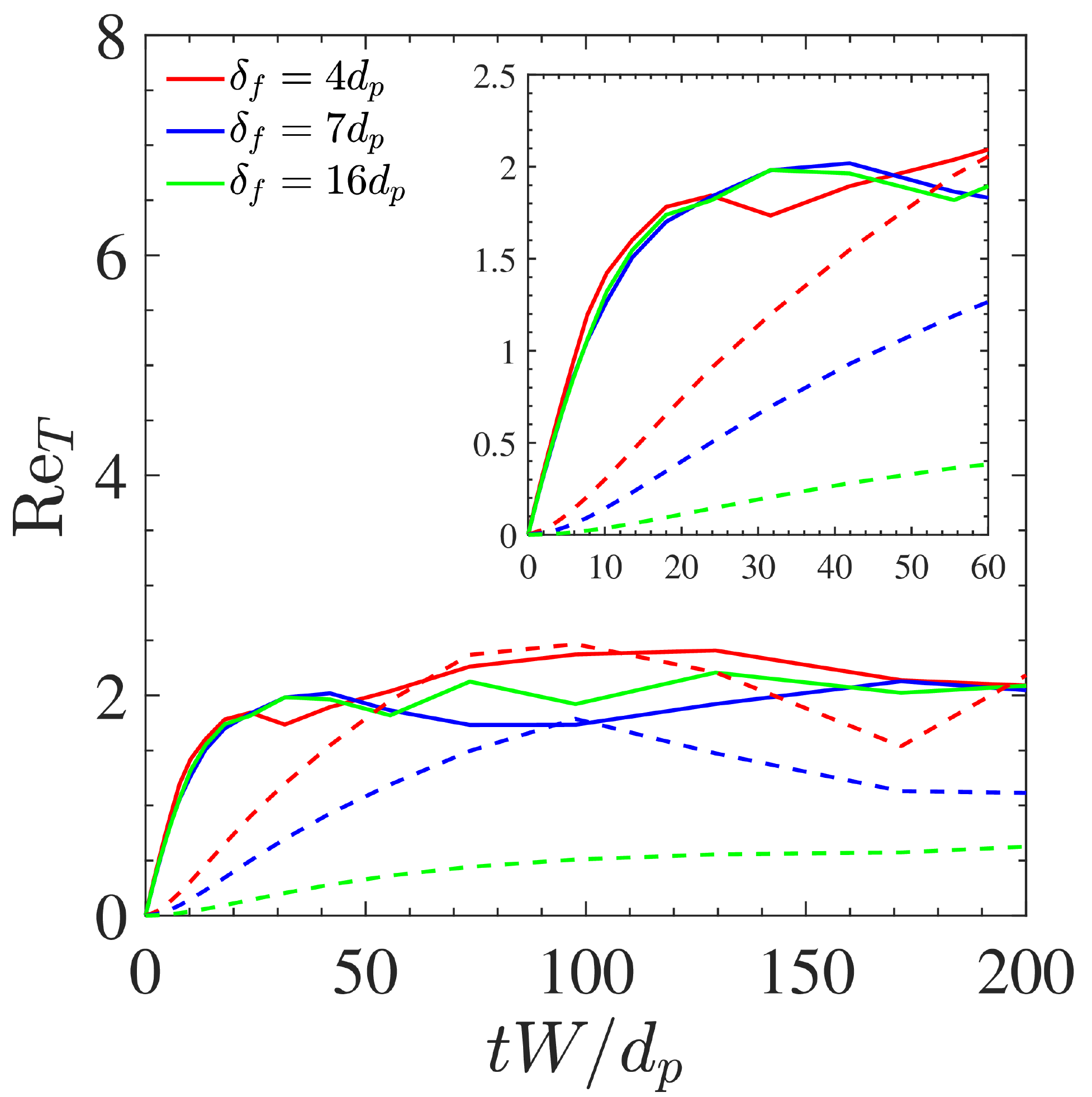}
        \caption{}
    \end{subfigure}
    \begin{subfigure}{0.35\textwidth}
        \centering
        \includegraphics[width=0.99\textwidth]{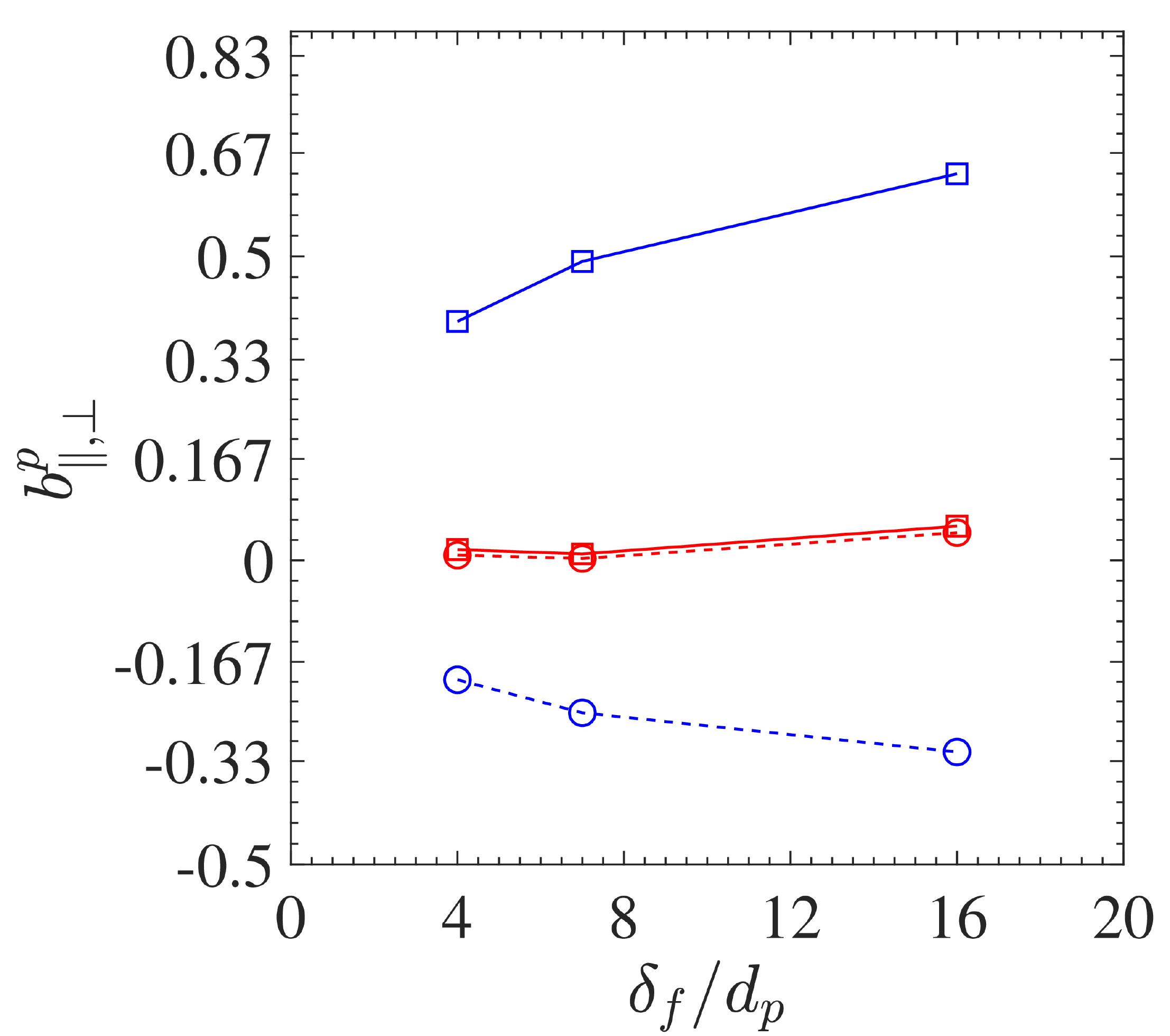}
        \caption{}
    \end{subfigure}
    \caption{\small{(a) ${\rm Re}_{T}$ evolution in the FHHS at ${\rm Re}_m =20; \hspace{0.5ex} \left\langle \phi \right\rangle =0.1; \hspace{0.5ex} \rho_p/\rho_f =100$ with varying $\delta_f$; solid lines denote stochastic EL and dashed lines denote standard EL. (b) Mean of the anisotropy tensor at steady state in the streamwise $b_{\parallel}^{p}$ (square) and transverse $b_{\perp}^{p}$ (circle) directions; red denotes stochastic EL while blue denotes standard EL.}}
\label{fig:Conv}
\end{figure}

We note that the FHHS has been examined in detail via PR--DNS with freely evolving particles by \citet{tang_direct_2016} and \citet{tenneti_stochastic_2016}. In these works it was shown that the FHHS is characterized by a rapid growth and sustainment of particle velocity variance. Additionally, \citet{tang_direct_2016} reported isotropic particle velocity fluctuations. Here, the anisotropy tensor $\bm{b}^{p}$ and thermal Reynolds number ${\rm Re}_{T}$ are examined at fixed simulation conditions ${\rm Re}_m =20; \hspace{0.5ex} \left\langle \phi \right\rangle =0.1; \hspace{0.5ex} \rho_p/\rho_f =100$ and varying filter size $\delta_f/d_p = 4, \hspace{0.5ex} 7, \hspace{0.5ex} 16$ (see Fig.~\ref{fig:Conv}). The new stochastic EL framework yields isotropic velocity fluctuations, ${\rm Re}_{T}$ curves that are are consistent with PR--DNS observations, and a steady state velocity variance that is relatively insensitive to $\delta_f$ refinement. By contrast, the standard EL framework yields highly anisotropic velocity fluctuations, biased to the streamwise direction, and results are a direct function of $\delta_f$. With the standard EL framework, the drag force statistics are \emph{inferred} from the resolved hydrodynamic fields; whereas the stochastic EL framework \emph{specifies} the drag force statistics through $\bm{F}_{d}^{(i)\prime\prime}$.

As prefaced at the end of Sec.~\ref{sec:Fdstats}, EL employs locally interpolated fluid quantities and instantaneous particle velocities to evaluate the hydrodynamic force acting on a particle. Consequently, force variance may be introduced into an EL simulation through the resolved fields. In fact, the dependence of velocity variance on filter width with the standard EL framework is a direct consequence of increased variation in resolved fields with decreasing $\delta_f$. To quantify the degree of force variation introduced by $\mathcal{V}_{p}^{(i)} \nabla \cdot \bm{\tau}$ and $\bm{F}_{d}^{(i)*}$, we compute the variance in these terms (denoted as $\sigma_{\rm un}^2$ and $\sigma_{\rm qs}^2$, respectively) at steady state and normalize them by the force variance obtained from Eq.~\eqref{eq:fvarcomp} at the mean conditions reported in Table ~\ref{tab:resvar}, $\left\langle \sigma_F^2 \right\rangle$. For the smallest filter width $\left(\delta_f/d_p = 4\right)$, variation in resolved fields leads to variation in quasi-steady drag that is comparable to $\left\langle \sigma_F^2 \right\rangle$; while the largest filter width $\left(\delta_f/d_p = 16\right)$ smears hydrodynamic fields to such a degree that negligible variance in quasi-steady drag occurs. Since fluctuating drag generates velocity variance independent of the resolved fields, $\sigma_{\rm qs}^2$ will not tend to zero as as $\delta_f \rightarrow \infty$ with the stochastic EL framework. Specifically, the particle velocity variance will feed back into the particle Reynolds number and $\sigma_{\rm qs}^2$ will tend to a constant as $\delta_f \rightarrow \infty$ with the stochastic EL framework. Future work that examines predictor-corrector methods for obtaining an exact force variance for abitrary filter width would be useful. However, we do not consider such a task here but note that the resolved force variances at $\delta_f/d_p = 7$ are of secondary significance in the homogeneous fluidization simulations.
\renewcommand{\arraystretch}{1.3}
\begin{table}
\caption{\small{Resolved force variances}}
\label{tab:resvar}
\begin{center}
\begin{tabular}{p{1.25cm} p{2cm}  p{1.5cm}  p{2cm}  p{1.5cm}}
\hline
\hline
 & \multicolumn{2}{c}{Standard EL} & \multicolumn{2}{c}{Stochastic EL} \\
 \cmidrule(lr){2-3}\cmidrule(lr){4-5}
$\delta_f/d_p$   & $\sigma_{\rm un}^2/\left\langle \sigma_F^2 \right\rangle$ & $\sigma_{\rm qs}^2/\left\langle \sigma_F^2 \right\rangle$  & $\sigma_{\rm un}^2/\left\langle \sigma_F^2 \right\rangle$ & $\sigma_{\rm qs}^2/\left\langle \sigma_F^2 \right\rangle$  \\
\hline
$16$ & $4.1 \times 10^{-5}$ & $0.002$ & $9.5 \times 10^{-6}$ & $0.163$\\
$7$  & $2.3 \times 10^{-4}$ & $0.127$ & $5.5 \times 10^{-5}$ & $0.210$ \\
$4$  & $1.3 \times 10^{-3}$ & $0.859$ & $3.9 \times 10^{-4}$ & $0.400$\\
\hline
\hline
\end{tabular}
\end{center}
\end{table}

\subsection{Fluidized homogeneous heating system (FHHS)} \label{subsec:FHHS}
\begin{figure}
    \centering
    \begin{subfigure}{0.32\textwidth}
        \centering
        \includegraphics[width=0.95\textwidth]{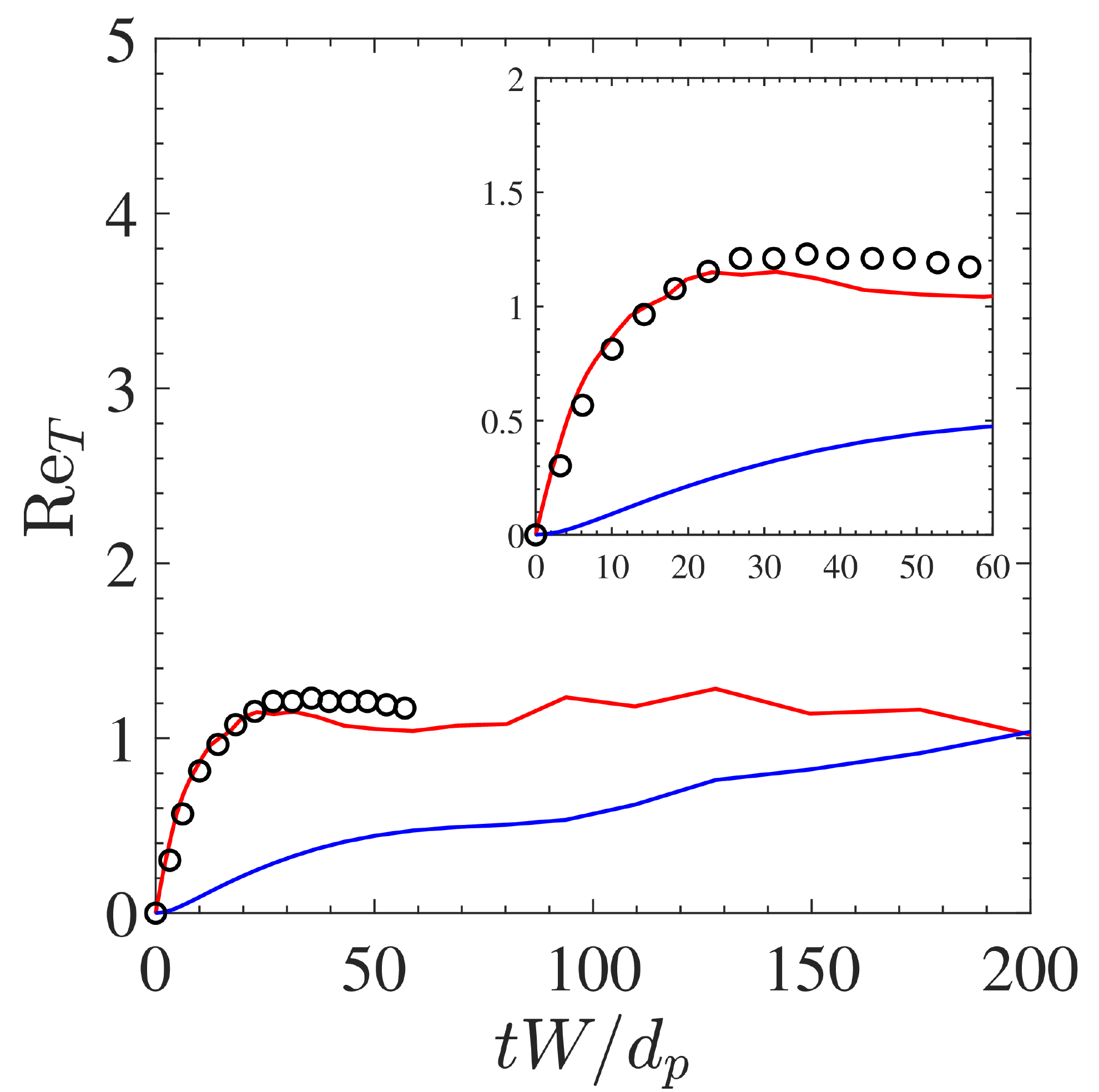}
        \caption{}
    \end{subfigure}
    \begin{subfigure}{0.32\textwidth}
        \centering
        \includegraphics[width=0.95\textwidth]{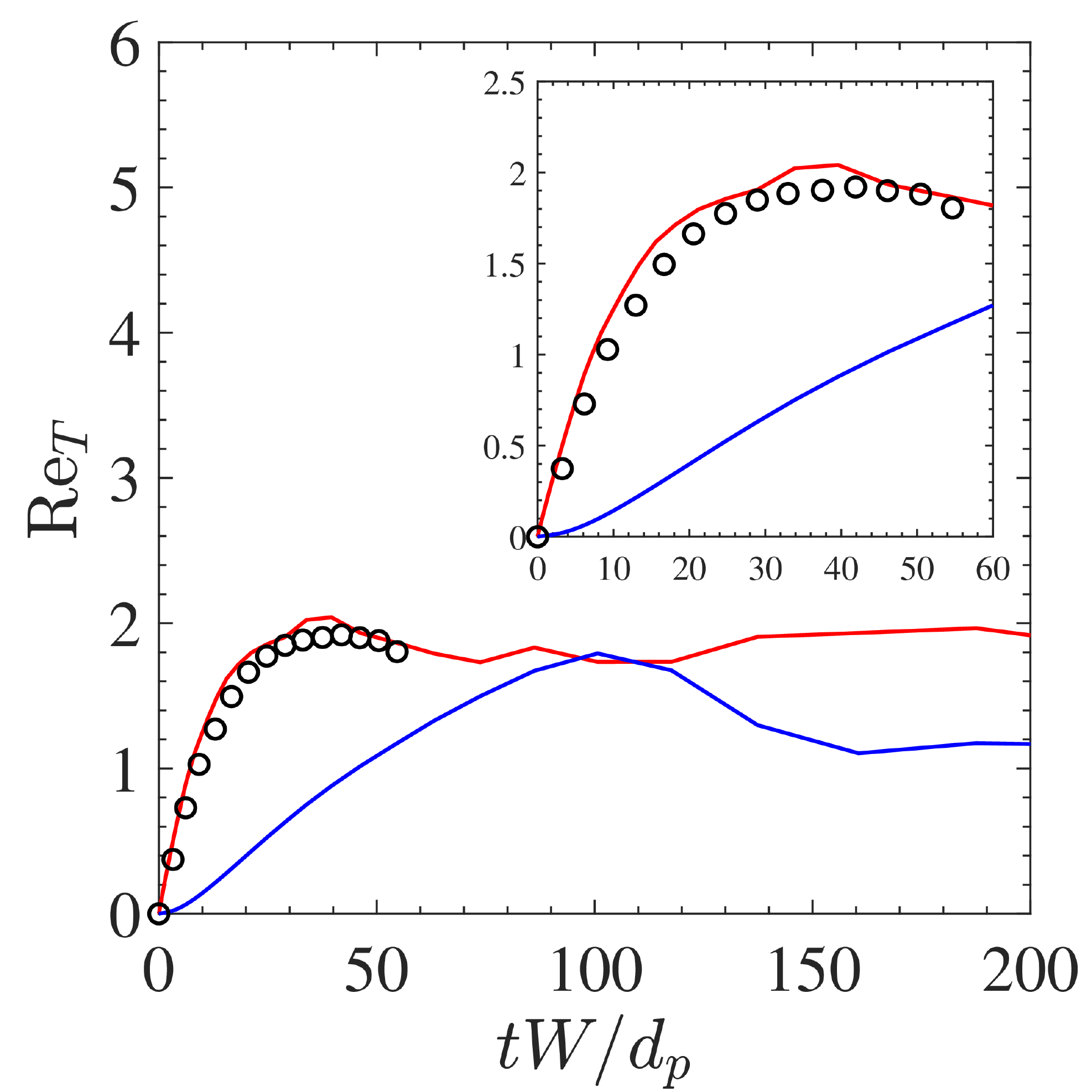}
        \caption{}
    \end{subfigure}
 \begin{subfigure}{0.32\textwidth}
        \centering
        \includegraphics[width=0.99\textwidth]{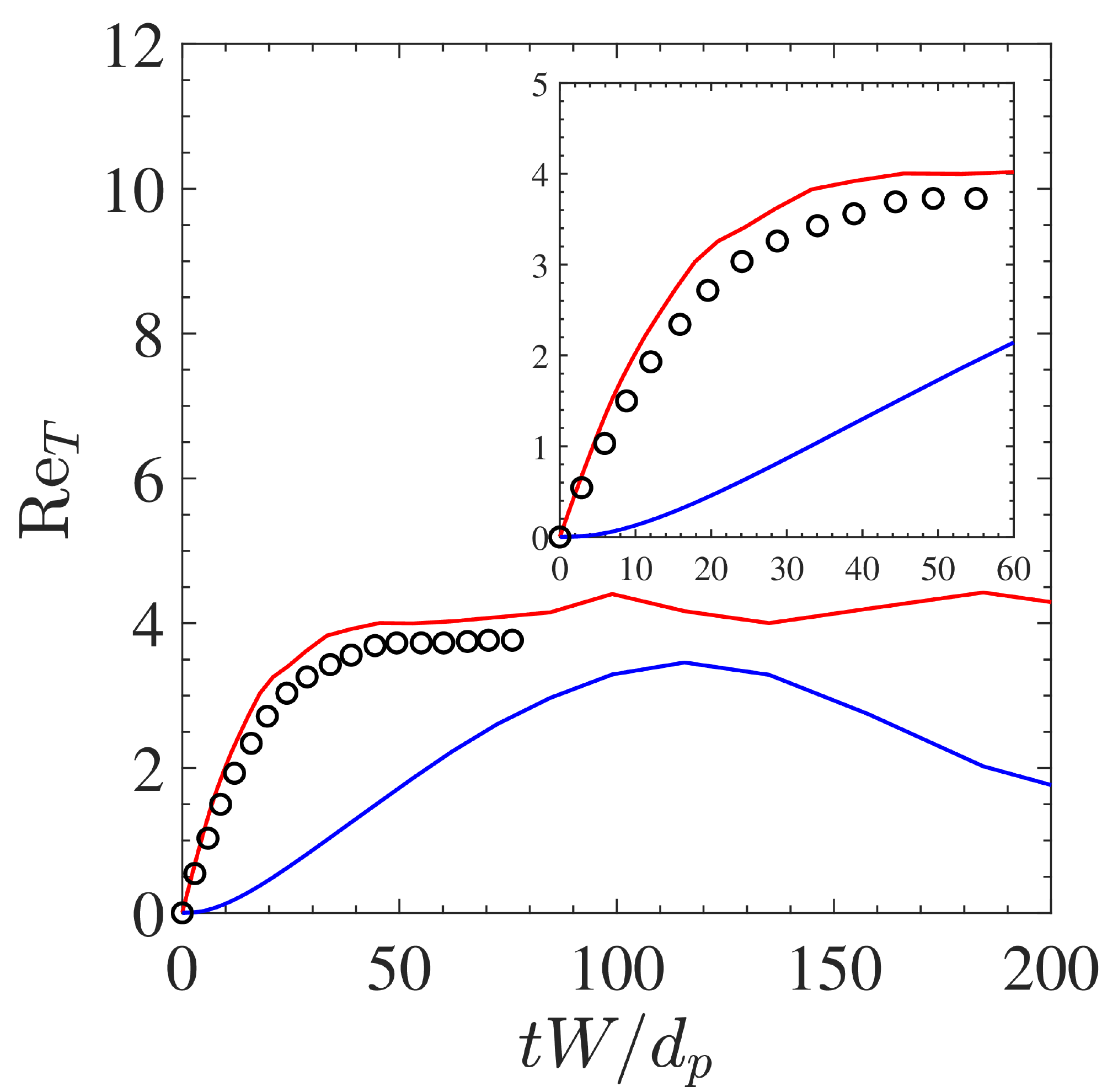}
        \caption{}
    \end{subfigure}
    \caption{\small{FHHS at (a) ${\rm Re}_m= 10$ (b) ${\rm Re}_m=20$ and (c) ${\rm Re}_m=50$. Stochastic EL (red lines), standard EL (blue lines), and PR--DNS results (black circles). All cases were simulated at $\left\langle \phi \right\rangle =0.1; \hspace{0.5ex} \rho_p/\rho_f =100$.}}
\label{fig:Resweep}
\end{figure}

\begin{figure}
    \centering
    \begin{subfigure}{0.35\textwidth}
        \centering
        \includegraphics[width=0.95\textwidth]{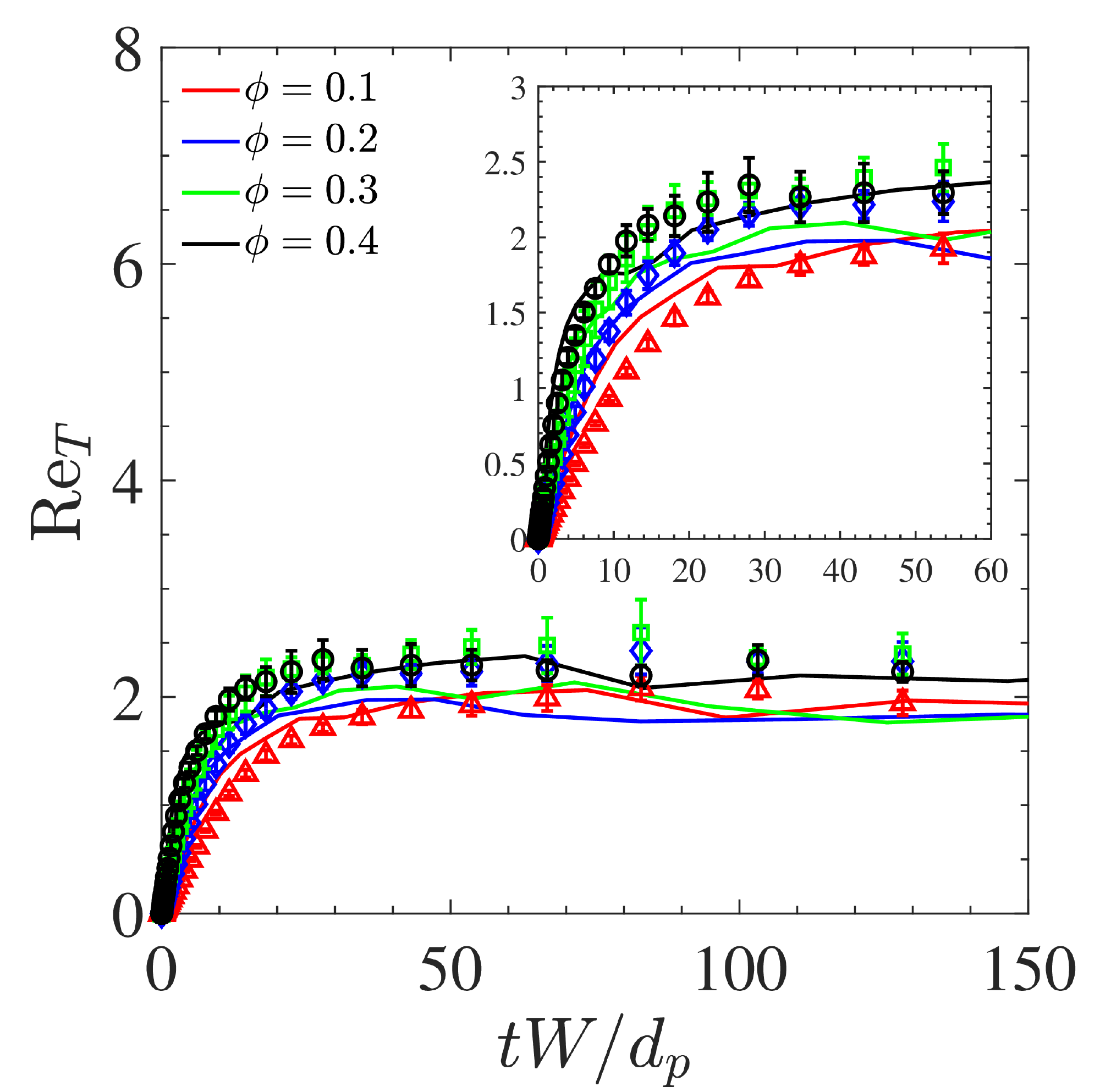}
        \caption{}
    \end{subfigure}
    \begin{subfigure}{0.35\textwidth}
        \centering
        \includegraphics[width=0.95\textwidth]{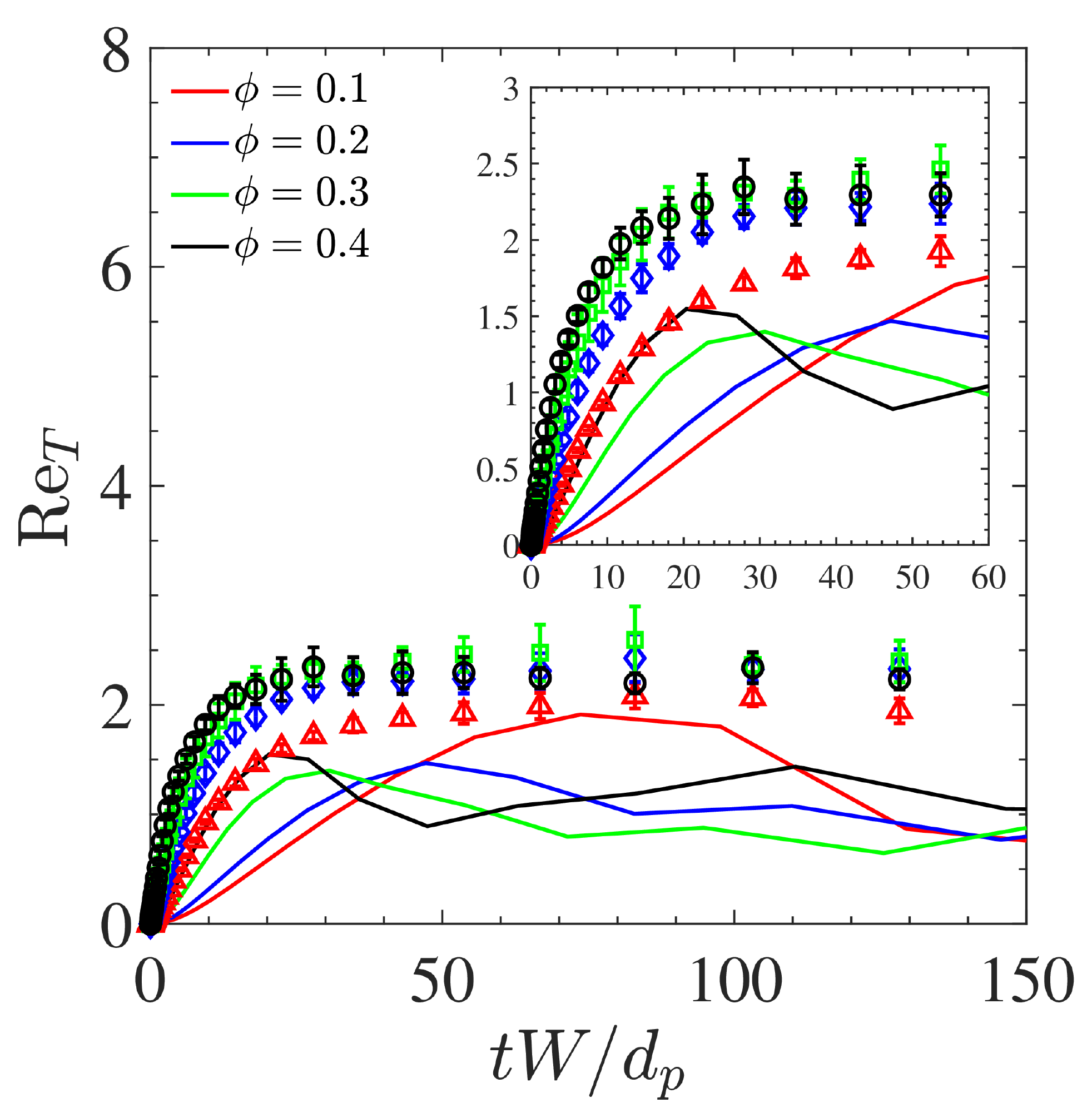}
        \caption{}
    \end{subfigure}
    \caption{\small{FHHS with (a) stochastic FL and (b) standard EL at $\left\langle \phi \right\rangle =0.1 , \hspace{0.5ex} 0.2, \hspace{0.5ex} 0.3, \hspace{0.5ex} 0.4$; ${\rm Re}_m= 20$; $\rho_p/\rho_f =100$. PR--DNS data (markers) and EL simulations (solid lines).}}
\label{fig:Phisweep}
\end{figure}

\begin{figure}
  \centering
    \begin{subfigure}{0.35\textwidth}
        \centering
        \includegraphics[width=0.95\textwidth]{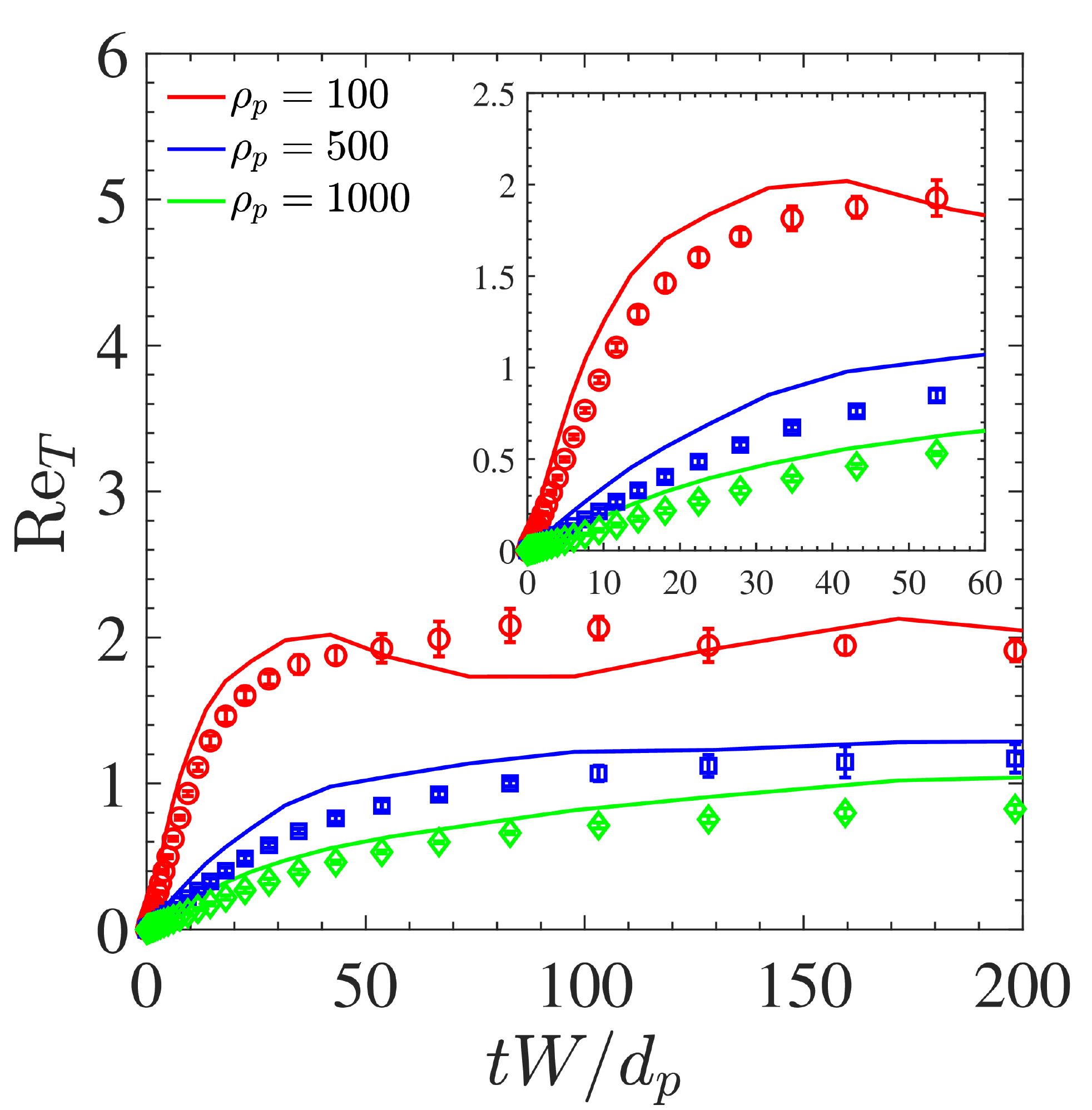}
        \caption{}
    \end{subfigure}
    \begin{subfigure}{0.35\textwidth}
        \centering
        \includegraphics[width=0.95\textwidth]{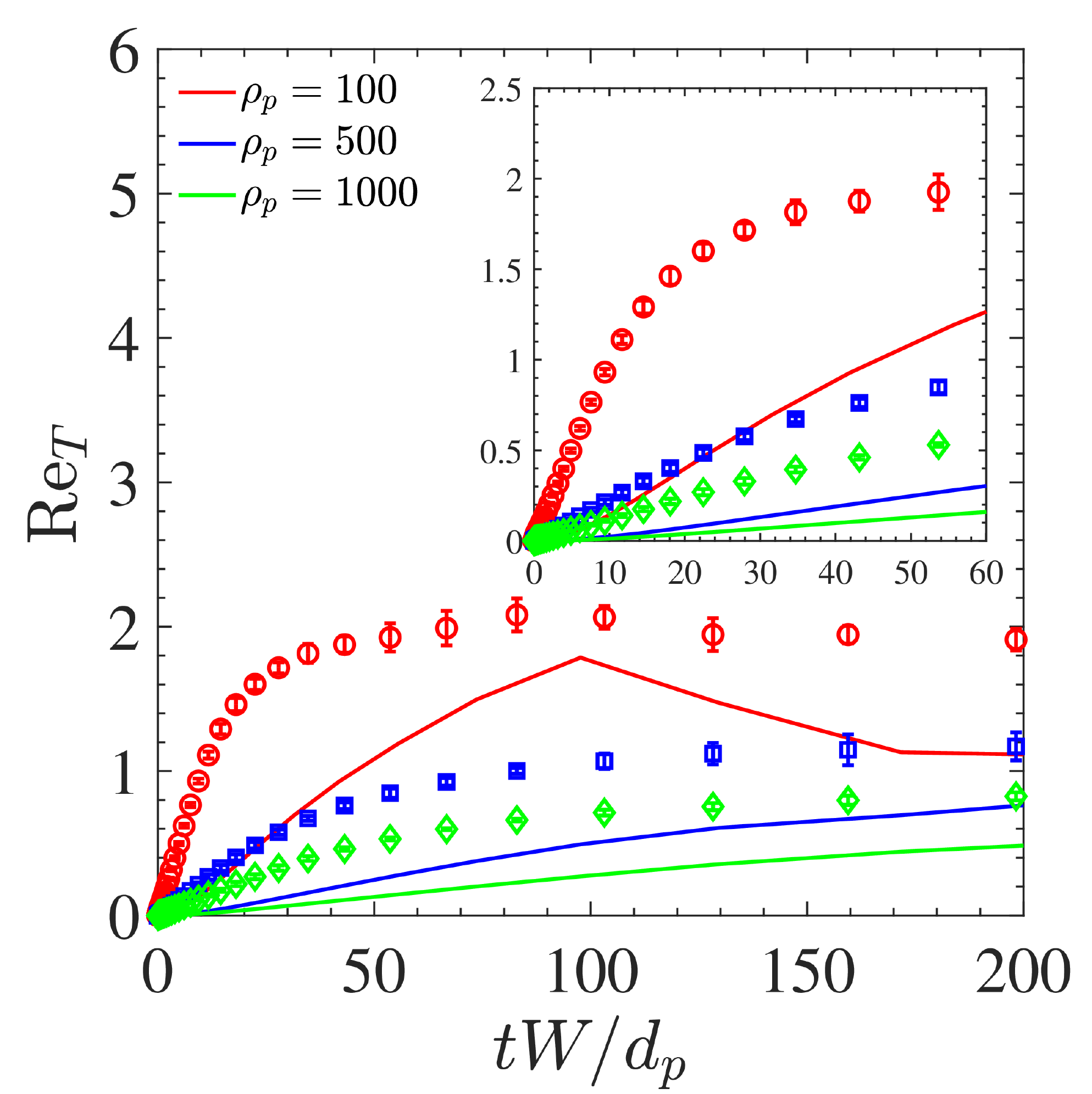}
        \caption{}
    \end{subfigure}
	\caption{\small FHHS with (a) stochastic FL and (b) standard EL at  $\rho_p/\rho_f=100, \hspace{0.5ex} 500, \hspace{0.5ex} 1000$; ${\rm Re}_m= 20$; $\left\langle \phi \right\rangle=0.1$. PR--DNS data (markers) and EL simulations (solid lines).}
\label{fig:Rhopsweep}
\end{figure}

In addition to the qualitative trends presented in Sec.~\ref{subsec:Con}, we draw direct comparison between PR--DNS, stochastic EL, and standard EL for the evolution of particle velocity variance in the FHHS. ${\rm Re}_{T}$ curves presented in \citet{tenneti_stochastic_2016} at varying ${\rm Re}_m$ and fixed $\left( \langle \phi \rangle; \, \rho_p/\rho_f \right)$ are employed as benchmark data for the comparisons illustrated in Fig.~\ref{fig:Resweep}. For the ${\rm Re}_m$ range considered in Fig.~\ref{fig:Resweep}, the stochastic EL framework is in strong agreement with PR--DNS results and captures not only the steady value of velocity variance but also the temporal growth. On the other hand, standard EL under-predicts the steady velocity variance and the temporal growth. Considering varying $\left\langle \phi \right\rangle$ and fixed $\left({\rm Re}_m; \, \rho_p/\rho_f\right)$, a comparison is drawn in Fig.~\ref{fig:Phisweep} to the PR--DNS data of \citet{tavanashad_effect_2019}. Similar trends are observed at varying solids volume fraction, where stochastic EL is in strong agreement with PR--DNS but the standard EL fails to build sufficient particle velocity variance. Error bars in Fig.~\ref{fig:Phisweep} correspond to 95\% confidence intervals computed over 5 realizations from the data of \citet{tavanashad_effect_2019}. We note that error bars are only provided here when comparing to the data of \citet{tavanashad_effect_2019}. Finally, we compare stochastic EL and standard EL to the PR--DNS data of \citet{tavanashad_effect_2019} at varying $\rho_p/\rho_f$ and fixed $\left({\rm Re}_m; \, \left\langle \phi \right\rangle \right)$ in Fig.~\ref{fig:Rhopsweep}. Combining the results obtained in Fig.~\ref{fig:Resweep}--\ref{fig:Rhopsweep}, it becomes apparent that standard EL neglects the role of sub-grid, neighbor-induced, drag fluctuations and consequently cannot capture the variance and dispersion obtained in homogeneous simulations. These results are consistent with the data provided in \citet{tenneti_direct_2010}, where quasi-steady drag was shown to act as a sink to granular temperature, causing standard EL frameworks to be overly dissipative in homogeneous fluidization. 

\subsection{Fluidized homogeneous cooling system (FHCS)} \label{subsec:FHCS}
\begin{figure}
    \centering
    \begin{subfigure}{0.32\textwidth}
        \centering
        \includegraphics[width=0.95\textwidth]{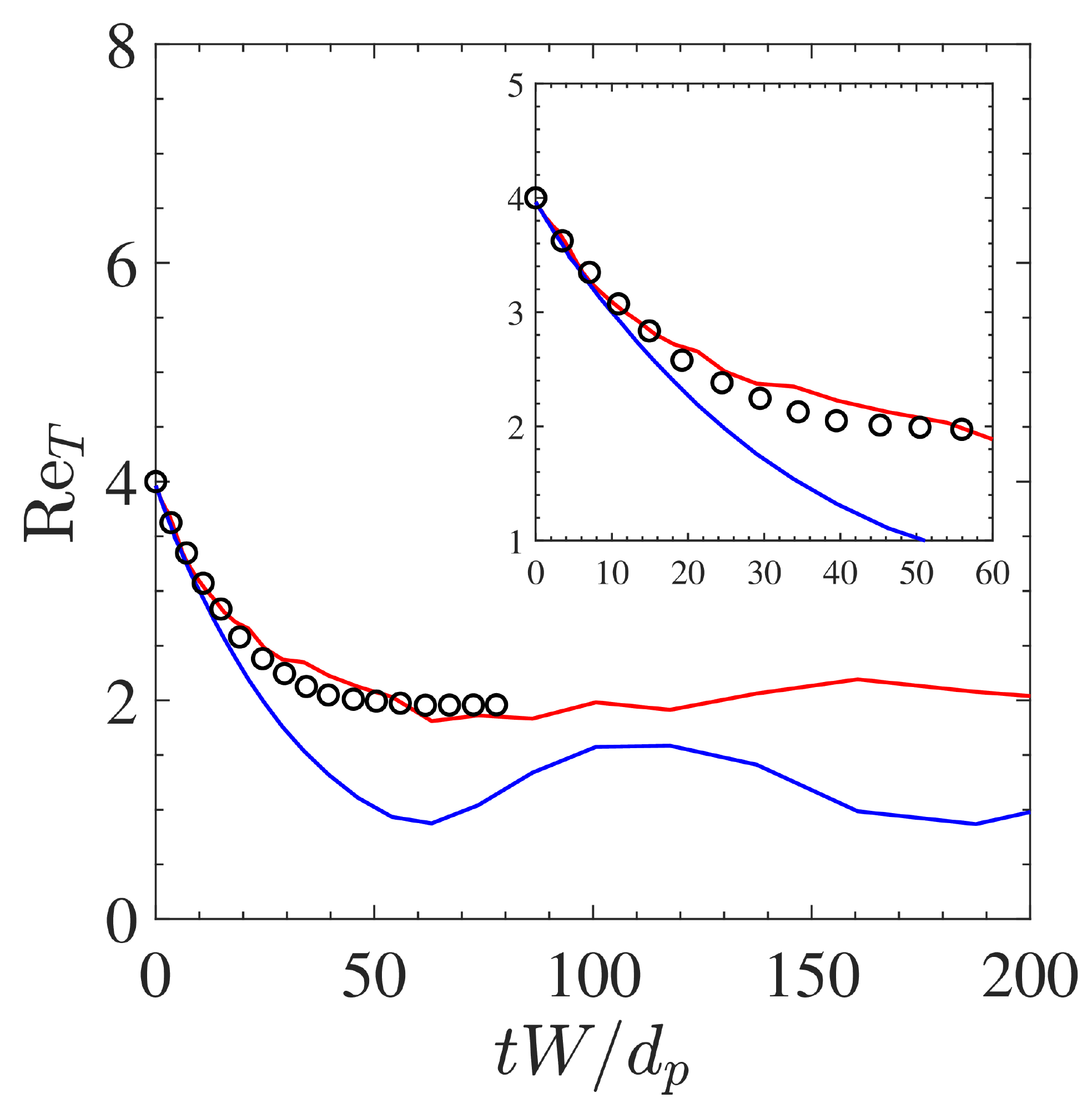}
        \caption{}
    \end{subfigure}
    \begin{subfigure}{0.32\textwidth}
        \centering
        \includegraphics[width=0.95\textwidth]{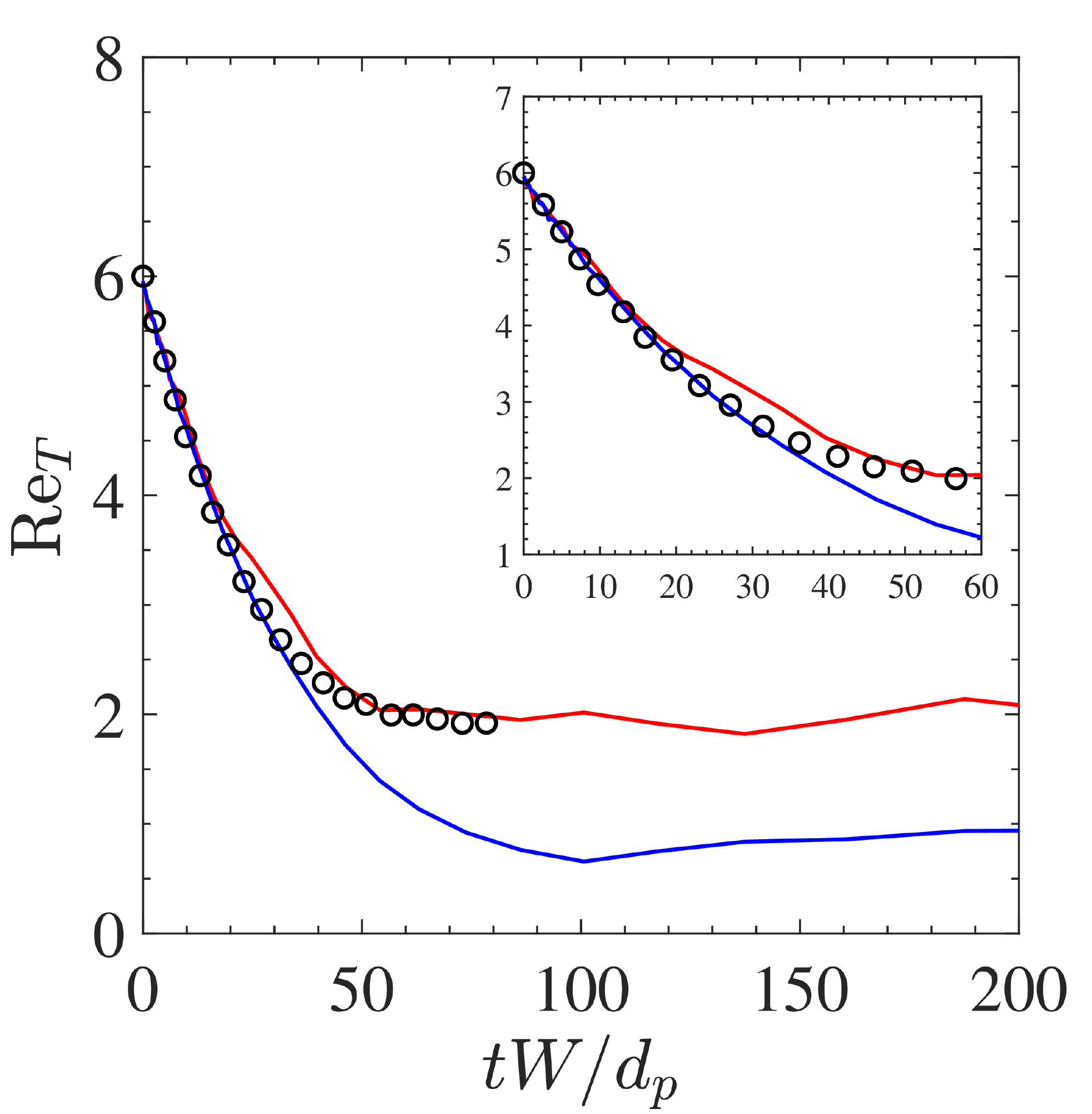}
        \caption{}
    \end{subfigure}
    \caption{\small{FHCS initialized with (a) ${\rm Re}_{T,0} = 4$ and (b) ${\rm Re}_{T,0}=6$. Stochastic EL (red lines), standard EL (blue lines), and PR-DNS results (black circles). All cases were simulated at ${\rm Re}_m= 20$; $\left\langle \phi \right\rangle =0.1$; $\rho_p/\rho_f=100$.}}
\label{fig:FHCS}
\end{figure}

For the FHHS system considered in Sec.~\ref{subsec:FHHS}, the suspension dynamics are dominated by fluid-mediated sources to particle velocity variance, since particles are initialized with zero velocity. By contrast, the FHCS system is initialized with an over-prescribed velocity variance and is thus dominated by hydrodynamic sinks to velocity variance. Directly extracting the fluid-mediated sources and sinks from PR--DNS show that the FHHS and FHCS stress these two terms, respectively \citep[see Fig. 8 of][]{tenneti_stochastic_2016}. As a result, the agreement between standard EL and PR--DNS is expected to improve for the FHCS since standard EL captures dissipation due to quasi-steady drag but not sources due to neighbor effects. Considering the ${\rm Re}_{T}$ curves from \citet{tenneti_stochastic_2016} as benchmark data, a comparison is drawn in Fig.~\ref{fig:FHCS} between PR--DNS, stochastic EL, and standard EL for the FHCS at fixed $\left({\rm Re}_m; \, \left\langle \phi \right\rangle; \, \rho_p/\rho_f \right)$ and varying initial condition ${\rm Re}_{T,0}$. In the FHCS, standard EL captures the temporal decay of velocity variance quite well but fails to sustain the correct velocity variance at steady state. By contrast, stochastic EL captures the temporal decay and steady velocity variance when compared to PR--DNS. These results highlight that the error observed with a standard EL method can depend upon the dynamics of the system, heating vs. cooling, in addition to the hydrodynamic regime considered.   

\section{Cluster-induced turbulence (CIT)} \label{sec:largescale}
Until this point, the domain size of the systems under consideration were small enough such that instabilities giving rise to heterogeneity in particle concentration were suppressed. However, large-scale flows with inertial particles will lead to the spontaneous formation of clusters due to two-way momentum coupling and/or dissipative collisions \citep{agrawal_role_2001,fullmer_clustering_2017,capecelatro_fluidparticle_2015}. As discussed by \citet{fullmer_clustering_2017}, neighbor-induced drag fluctuations act as a source of granular temperature that may contribute to cluster break up. To probe the role of neighbor-effects on cluster statistics, we consider simulations of fully-developed cluster-induced turbulence (CIT) with standard EL and stochastic EL. The CIT simulation geometry is essentially identical to the homogeneous fluidization described in Sec.~\ref{sec:verif}, with the exception that the domain length is increased to resolve the cluster length scale \citep{agrawal_role_2001}
\begin{equation}
\mathcal{L} = \tau_p^2 g_z.
\label{eq:Lcluster}
\end{equation}
Following \citet{capecelatro_effect_2016}, we set the streamwise domain length as $L_z = 32 \mathcal{L}$ to obtain converged statistics and avoid clusters falling in their own wakes. The simulation conditions are summarized in Table~\ref{tab:params3}. To quantify the degree of particle segregation in homogeneous isotropic turbulence, \citet{eaton_preferential_1994} proposed a clustering parameter $D$
\begin{equation}
D = \frac{\left\langle \left(\phi - \left\langle \phi \right\rangle \right)^2 \right\rangle^{1/2} - \sigma_p}{\left\langle \phi \right\rangle}, \label{eq:d}
\end{equation}
where $\sigma_p$ is the standard deviation in solids volume fraction for a random distribution of particles (evaluated at the beginning of the simulation for the initial random particle configuration). Here, we utilize $D$ as an indicator of the degree of clustering within the entire domain, so as to probe the effect of neighbor-induced drag fluctuations on cluster break up. While $D$ provides an estimate for the degree of clustering, we note that more detailed descriptions of heterogeneous media have been proposed that depend upon the viewing window size \cite{lu_local_1990,quintanilla_local_1997}. While these methods are beyond the scope of the present work, they do provide a more inclusive picture of the clustering spectrum, as opposed to a single value obtained with $D$. 

\begin{figure}
    \centering
 \begin{subfigure}{0.32\textwidth}
        \centering
        \includegraphics[width=0.99\textwidth]{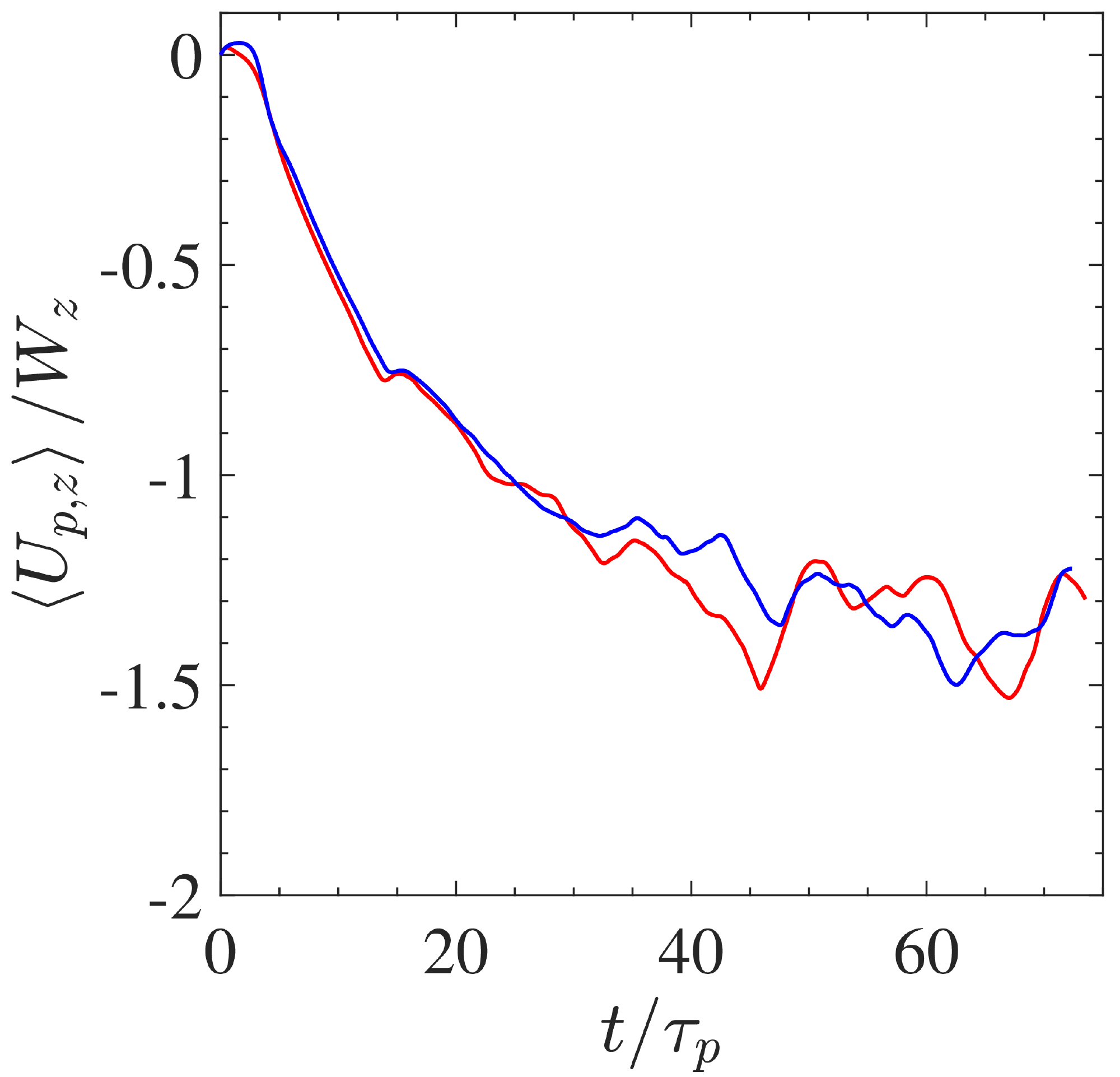}
        \caption{}
    \end{subfigure}
    \begin{subfigure}{0.32\textwidth}
        \centering
        \includegraphics[width=0.94\textwidth]{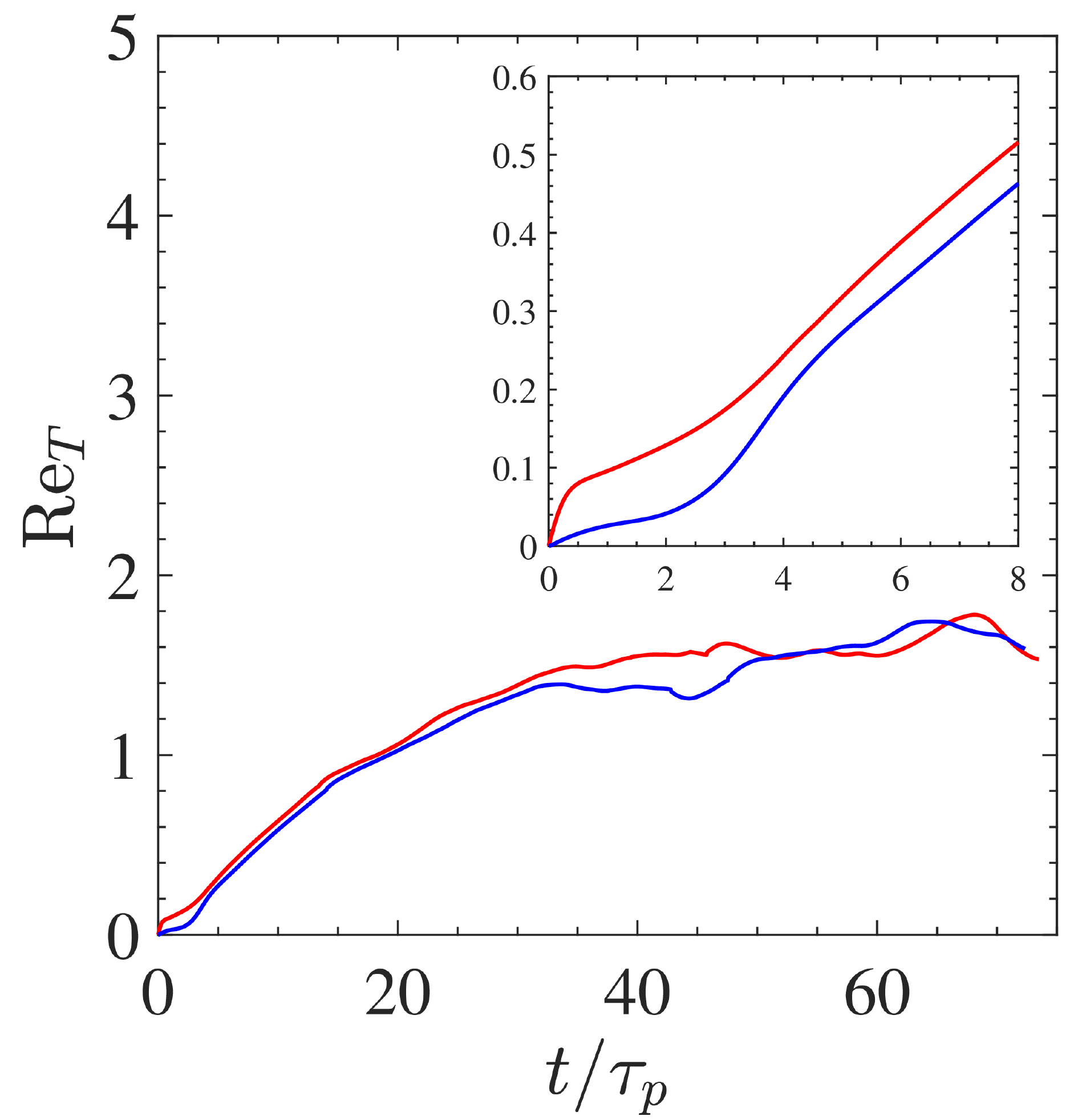}
        \caption{}
    \end{subfigure}
     \begin{subfigure}{0.32\textwidth}
        \centering
        \includegraphics[width=0.99\textwidth]{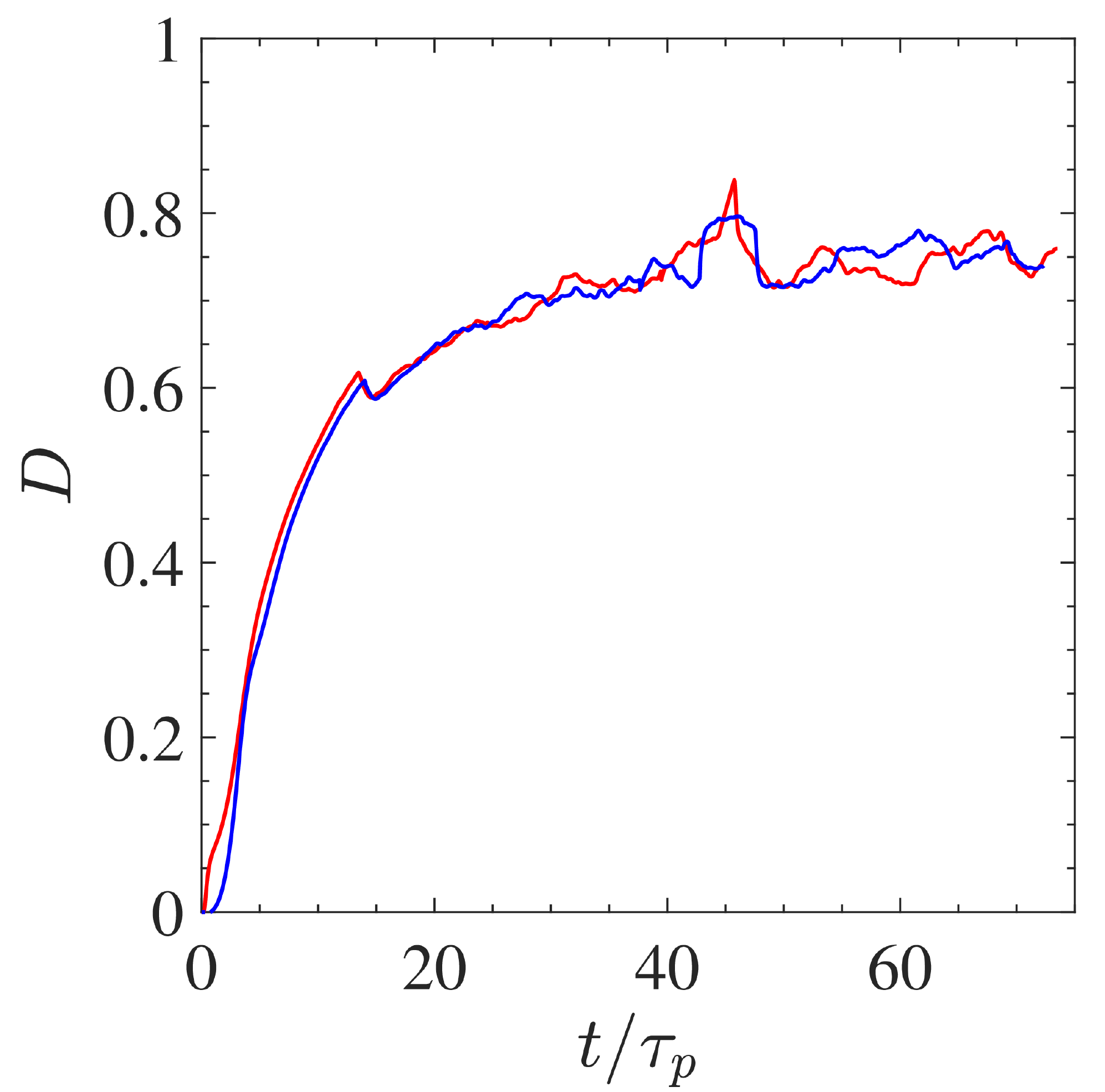}
        \caption{}
    \end{subfigure}
    \caption{\small{Evolution of the (a) normalized mean particle velocity, (b) particle velocity variance, and (c) clustering parameter in CIT. Blue is standard EL and red is stochastic EL.}}
\label{fig:CIT}
\end{figure}

Due to the presence of clusters, CIT exhibits drag reduction when compared to homogeneous systems. Specifically, entrainment of the surrounding fluid by particle clusters leads to a mean particle velocity in the streamwise direction $\left\langle U_{p,z}\right\rangle$ that is non-zero and in the direction of gravity (see FIG~\ref{fig:CIT} (a)). We note that simulations are completed here in a reference frame that moves with the mean slip velocity $W_z$, which is specified by the mean Reynolds number ${\rm Re}_m$ and the assumption that $\left\langle U_{p,z}\right\rangle \approx 0$; see discussion in Sec.~\ref{sec:verif}. Therefore, drag reduction in CIT leads to particle acceleration in the direction of gravity and mean settling velocities in the present simulations that are $\sim 2.25 W_z$. Examining the evolution of mean particle settling velocity over the course of the simulations shows there are negligible differences between stochastic EL and standard EL (see Fig.~\ref{fig:CIT} (a)). Similarly, negligible differences are observed between stochastic EL and standard EL for the clustering parameter (see Fig.~\ref{fig:CIT} (c)). Since the degree of clustering is tied to drag reduction, and by extension the mean settling velocity, $D$ and $\left\langle U_{p,z}\right\rangle$ are strongly correlated. For example, a significant reduction in $D$ would be reflected in an increase in $\left\langle U_{p,z}\right\rangle$, due to drag enhancement from homogenization of the particle phase. Since $D$ and $\left\langle U_{p,z}\right\rangle$ show consistent behavior, no appreciable change with the stochastic EL method, the model for neighbor-induced drag does not play a significant role in these domain averaged quantities, though it may alter the spectra obtained from the method of \citet{quintanilla_local_1997} .

\begin{figure}
    \centering
 \begin{subfigure}{0.4\textwidth}
        \centering
        \includegraphics[width=0.98\textwidth]{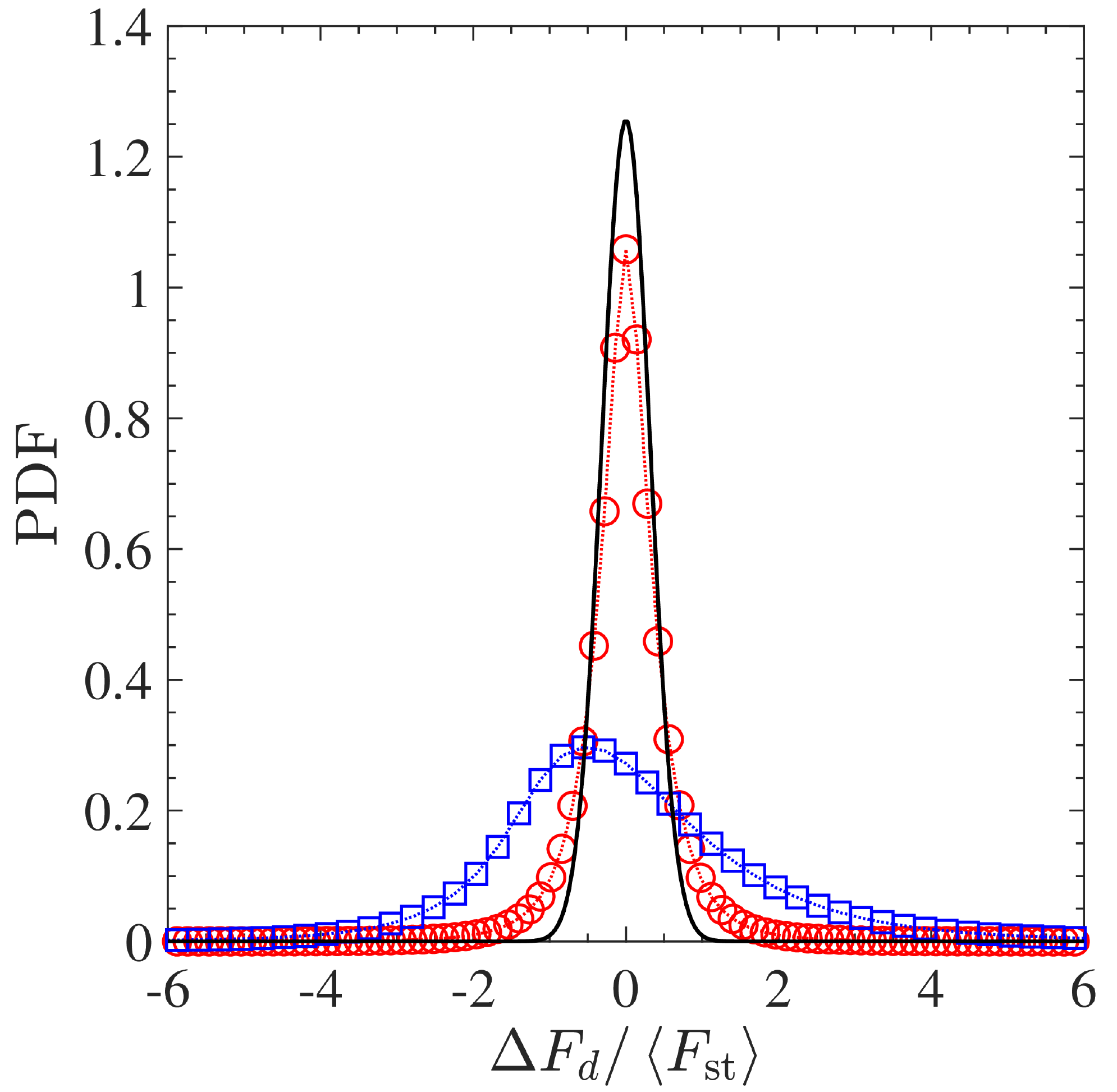}
        \caption{}
    \end{subfigure}
    \begin{subfigure}{0.4\textwidth}
        \centering
        \includegraphics[width=0.99\textwidth]{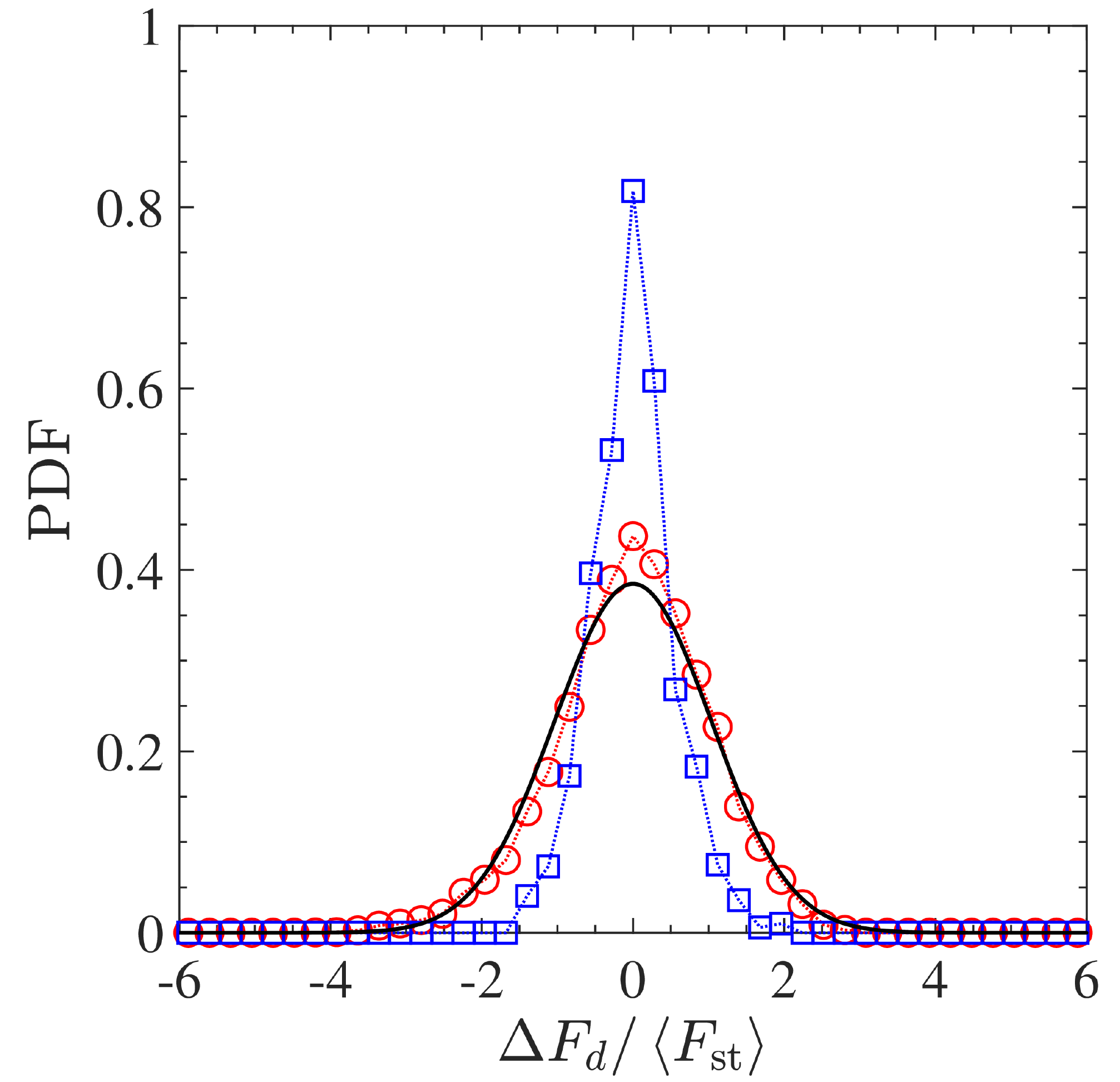}
        \caption{}
    \end{subfigure}
    \caption{\small{Probability distributions for (red) quasi-steady drag, $\bm{F}_d^{\prime \prime}$, and (blue) fluctuating drag, $\bm{F}_d^{*}$, in (a) CIT and (b) FHHS at ${\rm Re}_m= 20$; $\left\langle \phi \right\rangle =0.1$; $\rho_p/\rho_f=100$. $\Delta F_d$ denotes a removal of the mean and $\left\langle F_{\rm st} \right\rangle$ is Stokes drag evaluated at the mean conditions. The black line denotes $\mathcal{N}[0,~\left\langle \sigma_F \right\rangle]$, where $\left\langle \sigma_F \right\rangle$ is evaluated at the mean conditions.}}
\label{fig:CITPDF}
\end{figure}

Similar to mean settling velocity and clustering parameter, the particle velocity variance also shows minor deviation between stochastic and standard EL during transients and at steady state; see in Fig.~\ref{fig:CIT} (b). However, the beginning of the simulation $t/\tau_p \in \left[ 0 \hspace{0.5ex} 5\right]$ does show significant deviation between the velocity variance developed with stochastic EL and standard EL (see inset of Fig.~\ref{fig:CIT} (b)). At this point, particles are randomly, but still homogeneously, distributed within the domain. Under these conditions, the neighbor-induced drag fluctuation is the predominate source of velocity variance since clusters with sharp solids volume fraction interfaces have yet to form. As the system enters the transient stage $t/\tau_p \in \left[ 5 \hspace{0.5ex} 30\right]$, meso-scale particle structuring begins and the solids accelerate in the streamwise direction. The formation of clusters leads to significant quasi-steady drag variance at the edge of clusters, where large gradients in solids volume fraction and fluid velocity occur. These quasi-steady drag variances are resolved by a standard EL method and will facilitate granular temperature transport through the cluster via shear generation and collisional conduction. Therefore, heterogeneous systems provide additional sources to particle velocity variance that can dominate over the neighbor-effect. To illustrate this point, we extract probability distributions for quasi-steady and fluctuating drag in CIT and homogeneous fluidization (see Fig.~\ref{fig:CITPDF}). For the case of homogeneous fluidization, fluctuating drag is more significant than quasi-steady drag; while in the case of CIT, the exact opposite holds. In addition, we reiterate that the stochastic force is isotropic and was shown in Sec.~\ref{subsec:Con} to yield isotropic particle velocity fluctuations in homogeneous fluidization. However, for the case of CIT, stochastic EL and standard EL yield the same anisotropy $b^{p}_{\parallel} = 0.35$ and $b^{p}_{\perp} = -0.16$, which is further evidence that resolved quasi-steady drag is dominant over neighbor-induced drag.

\renewcommand{\arraystretch}{1.3}
\begin{table}
\caption{\small{Simulation conditions}}
\label{tab:params3}
\begin{center}
\begin{tabular}{p{1cm} p{2.5cm} p{1cm} p{0.75cm} }
\hline
\hline
$d_p$               & $90 \times 10^{-6}$ m & $\Delta x/d_p$       & 3 \\
$\mu_f$            & $1.0 \times 10^{-5}$ Pa$\cdot$s & $\delta_f/d_p$       & 7\\
$\rho_f$            & 1.0 kg/m$^3$ & $\mathcal{L}/d_p$ & 100\\
$\left\langle \phi \right\rangle$               & 0.05 & $L_x/\mathcal{L}$ & 8\\
${\rm Re}_m$   & 1 & $L_y/\mathcal{L}$ & 8\\
$\rho_p/\rho_f$ & 1000 & $L_z/\mathcal{L}$ & 32\\
\hline
\hline
\end{tabular}
\end{center}
\end{table}

\section{Conclusions} \label{sec:conclusion}
In the present study, we examine the role of higher-order drag statistics, originating from neighbor-induced fluid flow perturbations, in Eulerian--Lagrangian (EL) methods. A model was proposed for neighbor-induced hydrodynamic forces by treating the fluctuating drag as a stochastic variable that follows an Ornstein-Uhlenbeck process. Specifically, the force Langevin (FL) method detailed in \citet{lattanzi_stochastic_2020} was utilized here to construct a stochastic EL framework. Closures are provided for the theoretical inputs to the FL equation, integral time scale $\tau_F$ and force standard deviation $\sigma_F$, that are appropriate for inertial particles at moderate Reynolds numbers. Specifically, the integral time scale of the fluctuating drag is approximated with the mean-free time between successive collisions, derived from the kinetic theory of non-uniform gases. The standard deviation in drag force is closed with a new correlation based upon particle-resolved direct numerical simulation (PR--DNS) of fixed assemblies. The new stochastic EL framework specifies unresolved drag statistics through the stochastic force, leading to the correct evolution and sustainment of particle velocity variance when compared to PR--DNS of freely-evolving homogeneous suspensions. Since standard EL infers drag statistics from variations in the resolved flow, it cannot replicate the higher-order particle statistics (velocity variance and dispersion) observed in PR--DNS of homogeneous suspensions. Finally, the role of neighbor-induced drag fluctuations on cluster-induced-turbulence (CIT) is considered. In contrast to homogeneous fluidization, CIT is characterized by large gradients in solids volume fraction and fluid velocity. The aforementioned gradients provide a source for drag variance in standard EL, through the quasi-steady drag closure, that can dominate over neighbor effects. Since standard EL resolves the dominate modes for generating granular temperature in heterogeneous systems --- quasi-steady drag variation at cluster interface and collisional conduction --- we observe negligible change when employing the stochastic EL framework.

While emphasis is placed here on inertial particle suspensions at moderate solids loading and Reynolds numbers, we stress that the proposed methodology is a general framework that may be adapted to a variety of applications. Namely, the concept of higher-order drag statistics can be readily applied to higher-order statistics in Nusselt and Sherwood correlations, employed for simulation of heat and/or mass transfer. Additionally, compressible particle-laden flows often exhibit significant drag variation \emph{and} unsteady effects (added mass and Basset history). The present framework provides a stepping stone that may be leveraged by future work to account for such effects.  

\section*{Declaration of interests}
The authors report no conflict of interest.

\section*{Acknowledgements}
This material is based upon work supported by the National Science Foundation under grant no. CBET-1904742 and grant no. CBET-1438143. The authors acknowledge the Texas Advanced Computing Center (TACC) for providing Stampede2 compute resources under Extreme Science and Engineering Discovery Environment (XSEDE) grant no. TG-CTS200008.

\clearpage
\bibliography{main}

\end{document}